\documentclass[12pt,a4paper]{article}
\usepackage{jheppub}

\usepackage{color}
\usepackage{amsmath}

\usepackage{verbatim}   
\usepackage{subfigure}  
\usepackage{amsfonts}
\usepackage{amssymb}
\usepackage{mathrsfs}
\usepackage{graphicx}
\usepackage{slashed}
\usepackage{amsthm}

\usepackage{graphicx}
\usepackage{dcolumn}
\usepackage{bm}

\usepackage{color}
\usepackage{amsmath}

\usepackage{verbatim}   
\usepackage{subfigure}  
\usepackage{amsfonts}
\usepackage{amssymb}
\usepackage{mathrsfs}
\usepackage{graphicx}
\usepackage{slashed}
\usepackage{amsthm}

\def\bar#1{\overline{#1}}

\def\inv{^{\raise.15ex\hbox{${\scriptscriptstyle -}$}\kern-.05em 1}}
\def\lbar{{\lower.35ex\hbox{$\mathchar'26$}\mkern-10mu\lambda}} 

\let\p=\partial
\let\<=\langle
\let\>=\rangle

\let\+=\uparrow

\def\SSigma{\langle\Sigma\rangle}
\def\a{\alpha'}

\def\d{\textrm{d}}

\def\tr{\textrm{tr}}
\def\OO{\mathcal{O}}

\def\R{\mathcal{R}}

\newcommand{\beq}{\begin{equation}}
\newcommand{\eeq}{\end{equation}}

\newcommand{\bseq}{\begin{subequations}}
\newcommand{\eseq}{\end{subequations}}

\theoremstyle{definition}

\begin{document}
\preprint{\begin{tabular}{r} IPhT-T17/013 \\  \end{tabular}} 

\title{On Heterotic Vacua with Fermionic Expectation Values}
\date{\today}

\author[a]{Ruben Minasian}
\author[b]{Michela Petrini}
\author[b,c]{Eirik Eik Svanes}

\affiliation[a]{Institut de physique th\'eorique, Universit\'e Paris Saclay, CEA, CNRS, F-91191, \\Gif-sur-Yvette, France}
\affiliation[b]{Sorbonne Universit\'es, CNRS, LPTHE, UPMC Paris 06, UMR 7589, 75005 Paris, France}
\affiliation[c]{Sorbonne Universit\'es, Institut Lagrange de Paris, 98 bis Bd Arago, 75014 Paris, France}

\emailAdd{ruben.minasian@cea.fr}
\emailAdd{petrini@lpthe.jussieu.fr}
\emailAdd{esvanes@lpthe.jussieu.fr}

\null\vskip10pt

\date{\today}

\abstract{We study heterotic backgrounds with non-trivial $H$-flux and  non-vanishing expectation values of fermionic bilinears, often referred to as gaugino condensates. The gaugini appear in the low energy action via the gauge-invariant three-form bilinear $\Sigma_{MNP}=\tr\:\bar\chi\Gamma_{MNP}\chi$. For Calabi-Yau compactifications to four dimensions, the gaugino  condensate corresponds to  an internal three-form $\Sigma_{mnp}$ that  must be a singlet of the holonomy group.  This condition does not hold anymore when an internal $H$-flux is turned on and  $\OO(\a)$ effects are included. In this paper we study flux compactifications to three and four-dimensions on $G$-structure manifolds. We derive the generic conditions for supersymmetric solutions. We use integrability conditions and  Lichnerowicz type arguments to  derive a set of constraints whose solution, together with supersymmetry, is sufficient for finding backgrounds with gaugino condensate.}

\keywords{Heterotic Supergravity, Supersymmetry, Gaugino Condensates}

\maketitle

\section{Introduction}
The study of fermion, notably gaugino, condensates in heterotic string  was primarily motivated by  attempts to break supersymmetry  while preserving a zero cosmological constant at tree  level \cite{Derendinger198565, Dine:1985rz, DERENDINGER1986365}, based on earlier work in supergravity \cite{NILLES1982193, Duff198337, Ferrara1983457}. More recently fermionic condensates have also been considered in the context of supersymmetric compactifications, often in association with  a non-trivial $H$-flux (see e.g. \cite{LopesCardoso:2003sp, Frey:2005zz, Manousselis:2005xa, Derendinger:2006hr, Lowen:2008xh, Held:2010az, Lechtenfeld:2010dr, Chatzistavrakidis:2012qb, Klaput:2012vv, Gemmer:2013ica, Quigley:2015jia}).  In the supersymmetry transformations and the low energy action only the gaugino bilinear  with three gamma matrices 
\begin{equation}
\Sigma_{MNP}  =  \tr\:\bar\chi\Gamma_{MNP}\chi \,  , 
\end{equation}
appears \cite{Bergshoeff:1989de}. At a formal level,  in order to preserve supersymmetry at order $\a$ (and eventually satisfy the bosonic equations  of motion),  an interplay between the three-form $\Sigma \, $, the NS three-from flux $H$ and the  geometric data of the internal space is required. 

Poincar\'e invariance requires the  vev's of the individual fermions  to vanish, which does not need to be the case for   vev's of fermion bilinears such as $\Sigma$. As far as Poincar\'e invariance goes, one can roughly impose on $\Sigma$ the same type of conditions as for  fluxes.
This has very restrictive implications for compactifications to four dimensions, where 
only the components of the gaugino bilinear in the internal six-dimensional space 
are allowed.
 Moreover, if the internal manifold has  $SU(3)$ holonomy, i.e. a Calabi-Yau space,
the gaugino condensate must be a singlet of the  holonomy group. This is due to the fact that the  component
 of $\chi$ that is massless in four dimensions is an $ SU(3)$ singlet \cite{Dine:1985rz}.  This forces the internal three-form to be proportional to the holomorphic top-form:
\begin{equation}
\<\Sigma \, \>= \Lambda \Omega + c.c. \:,
\end{equation}
where $\Lambda \sim \<\tr\:\bar\chi^{(4)} (1+ \gamma_5) \chi^{(4)} \>$ can be viewed as the ``four-dimensional condensate". When compactifying to three dimensions, similar arguments can be made when the internal manifold is a compact space of $G_2$ holonomy. Now the condensate is forced to be proportional to the associative three-form $\varphi$. However, an external ``spacetime filling" component of $\Sigma$ is allowed by  Poincar\'e invariance  and has to be turned on \cite{Gemmer:2013ica}.  

In this paper we are interested in  heterotic string backgrounds  that involve non-trivial gaugino condensates 
and a non-trivial  $H$-flux.  
Turning on the  $H$-flux necessitates the inclusion of $\OO(\a)$ effects and the consideration of internal spaces that are not Ricci-flat, and hence are not of special holonomy. As it is usual in flux compactifications, one is interested in manifolds that support nowhere vanishing (not necessarily covariantly constant) spinors, or equivalently a  so-called $G$-structure.  
We will consider  compactifications to four-dimensions on $SU(3)$ structure manifolds and to  three-dimensional $G_2$ structure manifolds.
Considering non compact  internal seven-manifolds that are foliations over  suitably chosen six-dimensional compact spaces allows to describe  also
four-dimensional domain-wall backgrounds.  
Importantly for our purposes, when studying the lower-dimensional effective theories on such manifolds, one is no
longer obliged to restrict  to a strictly massless lower-dimensional spectrum.  As a consequence the gaugino condensate is no longer constrained to be a singlet of the structure group.
Clearly we still have to require that,  upon inclusion of $\SSigma$ and $\OO(\a)$ effects, the ten-dimensional equations of motion and Bianchi identities are all satisfied.

Our strategy will be to start from ten-dimensional equations, and include nontrivial condensate $\< \Sigma \>$.\footnote{When this does not lead to confusion, we shall drop the brackets  $\< \,\, \>$ when talking about  vev's and just use $\Sigma$ for the components of the condensate. When talking about $\Sigma$, terms as  ``gaugino condensate" and ``gaugino bilinear" will be used interchangeably. }   This may appear to be as bad as having fermion vevs, since the ten-dimensional Poincar\'e invariance is broken. However we shall then restrict ourselves to backgrounds with three- and four-dimensional Poincar\'e invariance. Fermion vevs will still not be allowed but some components of $\Sigma$ are compatible with the symmetry. We remark that there are bi-spinor contributions to Einstein and dilaton equations of motion, of the form $\sim \alpha' \, \< \chi^\dagger \gamma_{(M} \nabla_{N)} \chi \>$ and $\sim \alpha' \, \<\chi^\dagger\slashed\nabla\chi \> $ respectively. We shall assume that only flux-like objects, i.e. $\Sigma$, can have nontrivial expectation values, and hence from now on we can ignore the gaugino kinetic terms.

As already mentioned, for backgrounds of the the form $M_{10} = M_3 \times X_7$ 
 Poincar\'e invariance is compatible with non-zero external components  for 
both $\Sigma$ and the $H$-flux.\footnote{ Note that since  $\Sigma$ is a bilinear of ten-dimensional spinors, both its internal and external components will be products of internal and external spinor bilinears. } 
Note that we only need to require that the internal manifold is spin, as this automatically leads to existence of nowhere vanishing spinors and hence $G_2$ structures.\footnote{In fact, every spin seven-manifold has an $S^3$ of $G_2$ structures, or equivalently an $SU(2)$ structure. This allows to decompose the ten-dimensional spinors into twisted products of external and internal spinors, and eventually may lead to solutions with higher amount of preserved supersymmetry, than allowed by the direct product spinorial ansatz (or  preservation of supersymmetry in cases where the direct product ansatz of internal and external spinors cannot preserve any). It may be of interest to study these cases further, however they stay outside the scope of our paper.
As far as the geometric structures go, we shall make use of a single $G_2$ structure on the internal seven-manifold and hence employ the direct product anstaz.} In spite of  the formal similarity  with  the associative three-form $\varphi$, the condensate  $\Sigma$ does not define an alternative $G_2$ structure, as it may vanish pointwise. As we shall see in section \ref{sec:split37},  
the conditions imposed by supersymmetry imply the $H$ equation of motion and  are somewhat under-constrained. 
For $M_{10} = M_4 \times X_6$ only internal components of $H$ and $\Sigma$ are allowed. Even in this case the  conditions imposed by supersymmetry
are under-constrained.

Without condensate, by a generalisation of Lichnerowicz formula, the bosonic action can be written as \cite{Coimbra:2014qaa}
\begin{align}
\label{eq:nocondBismut2***}
S_b =  BPS^2 =   & 4   \int_{M_{10}}e^{-2\phi} [(\slashed D^0\epsilon)^\dagger\slashed D^0\epsilon-(D^0_M\epsilon)^\dagger D^{0 \, M}\epsilon \nonumber \\ 
& +\frac{\a}{96}\left(\tr\,\epsilon^\dagger\slashed F\slashed F\epsilon-\tr\,\epsilon^\dagger\slashed R^-\slashed R^-\epsilon\right)] +\OO(\a^2)\:,
\end{align}
where we took   the ten-dimensional supersymmetry parameter  $\epsilon$ to have norm one  ($\epsilon^\dagger\epsilon=1$),
and  $D^0$ and $D^0_M$ are differential operators appearing in the (modified) dilatino and gravitino variations  \eqref{eq:varmoddil} and 
\eqref{eq:varpsi}.  The  superscript $0$ denotes the absence of condensate $\Sigma$.  
Note that the equality  \eqref{eq:nocondBismut2***}
 implies integration by parts  and vanishing boundary conditions for the fields.  Moreover \eqref{eq:nocondBismut2***} reproduces the heterotic action only after imposing  Bianchi identity for $H$ 
  \begin{equation}
\d H - \frac{\a}{4} \left( \tr (R^-)^2 - \tr F^2 \right) = 0 \, .
\end{equation}
From \eqref{eq:nocondBismut2***} it follows that the action vanishes for ten-dimensional supersymmetric solutions. A variation of \eqref{eq:nocondBismut2***} further implies that supersymmetric solutions also solve  the equations of motion, provided the Bianchi identity is satisfied.\footnote{Note that  supersymmetry (see the variation of the covariant derivative of the gravitino) implies  that the curvature of the torsionful connection $R^-$ satisfies the Hermitean Yang-Mills equation to the appropriate order in $\a$.}

While there are reasons to believe that generic supersymmetric theories should satisfy this kinds of generalised Lichnerowicz theorem, this clearly applies only to the bosonic action (indeed $D^0$ and $D^0_M$ appear in the fermionic variations, and there are no analogues for the bosonic ones). So the condensate violates the  $BPS^2$ property, and solving supersymmetry conditions (and the Bianchi identity) no longer guarantees solving equations of motion.
Yet we shall show that, under the assumption that one can integrate by parts  the Lichnerowicz formula  can be generalised  to cast the action in the form\footnote{
We  formally treat  the condensate as three-form in ten dimensions $\Sigma$. Of course, we should be aware that $\Sigma$ is breaking the ten-dimensional Poincar\'e invariance.}
\beq
\label{BPSmodintro}
S_b = BPS^2 + \Delta S(\Sigma) \, .
\eeq
This  simplifies the analysis of the equations of motion significantly. Indeed  the equations of motion for the  dilaton, $B$-field and metric 
derived from \eqref{BPSmodintro} have the supersymmetry constrains already taken into account. On a more formal level, this exercise may also turn out to be useful in understanding the limits of application of the generalised Lichnerowicz theorem to supersymmetric theories.

As mentioned above,  the derivation of \eqref{eq:nocondBismut2***} involves integration by parts and hence assumes the  vanishing of the fields at infinity. This will not be true for $AdS$ backgrounds or for domain walls.   For these cases  we derive the integrability conditions for  the supersymmetry variations corrected by the fermion bilinears.  The result are equations of the form
\beq
BPS^2 = E.O.M + f(\Sigma)
\eeq
where  $ f(\Sigma)$ is a function of  the gaugino condensate.  The l.h.s of the above  equation vanishes due to supersymmetry and the equations 
of motion are implied if $\Sigma$ satisfies the constraints  $ f(\Sigma)=0$. One can show that when the internal space is compact or when the boundary conditions at infinity allow for integration by parts the two types of analysis agree. 

The integrability conditions can also be reformulated in a way that  allows 
to  discuss  non-supersymmetric solutions and involves a curios phenomenon
labeled as ``fake supersymmetry".  The crucial observation is that the equations of motion  with non-trivial condensate have the same 
form as with zero condensate but with  the replacement 
\beq
H\rightarrow\bar H = H + \frac{\a}{2} \Sigma \, . 
\eeq
One can define ``fake supersymmetry" equations,   $BPS(\bar{H})$, that have the same form as the BPS equations with
zero condensate but with   $H$ replaced by $\bar{H}$.  Squaring them we find 
\beq
BPS^2(\bar{H})  = E.O.M   \,. 
\eeq
Then solutions of the equations of motion  can be found by solving the fake BPS conditions. The solutions will not be supersymmetric.
Since the fake BPS conditions are linear equations, this formulation can be a useful tool to study supersymmetry breaking by a gaugino condensate, as was the original idea of \cite{Derendinger198565, Dine:1985rz, DERENDINGER1986365}.

The structure of the paper is as follows.  In section \ref{sec:hetsugra} we give the supersymmetry conditions of heterotic theory with non-zero
gaugino condensate and we recall Lichnerowicz formula. Then we discuss the integrability conditions and the generalisation of 
 Lichnerowicz argument.  We end this section with a   discussion of  the fake supersymmetry conditions. 
Sections \ref{sec:split46} and \ref{sec:split37} are devoted to the analysis of the  general supersymmetry conditions for 
compactifications to four and three dimensions on manifolds of $SU(3)$ and $G_2$ structure, respectively.  The latter also include domain walls
in four dimensions. 
An explicit example of  domain wall solution is presented in section \ref{sec:Ex}.  Some background and technical material can be found in the appendices.

\section{Heterotic Supergravity with non trivial gaugino bilinears}
\label{sec:hetsugra}

The field content of   $E_8\times E_8$  heterotic supergravity consists of the metric,  the NS two-form $B$,  the dilaton $\phi$, the gauge-field $A$,
plus  the gravitino $\psi_M$, the dilatino $\lambda$ and the gaugino $\chi$.  The bosonic action and equations of motion, as well
as the supersymmetry variations, are given in appendix \ref{app:hetsugra}. 

In this paper we are interested  in solutions of  the theory with  non-trivial fermionic bilinears.
 Specifically, we consider what happens if we allow for gaugino condensation. This means that some  components of  the three-form
\begin{equation}
\Sigma_{MNP}= \tr\:\bar\chi\Gamma_{MNP}\chi  \, 
\end{equation}
can take non-zero values.

The equations of motion and supersymmetry conditions with non-zero fermionic bilinears have been derived in 
\cite{Bergshoeff:1989de}. Keeping only the gaugino bilinears they read\footnote{Note that, in order to have a well-defined theory at first order in $\a$,
 the gaugino supersymmetry variation need only be specified to zero-th order. This is due to the fact that the gauge fields of heterotic supergravity are already an $\OO(\a)$ effect.}
\bseq
\begin{align}
\label{eq:varpsi}
\delta\psi_M=\:&  \: D_M \epsilon = \nabla_M\epsilon+\frac{1}{4} H_{M} \epsilon +\frac{\a}{16} \slashed \Sigma \Gamma_M \epsilon+\OO(\a^2) \, , \\
\label{eq:vardil}
\delta\lambda=\:&  -\frac{\sqrt{2}}{4}\,\mathcal{P}\epsilon = -\frac{\sqrt{2}}{4}\Big[\slashed\p\phi+\frac{1}{2}\slashed H-\frac{\a}{8} \slashed \Sigma \Big]\epsilon+\OO(\a^2) \, , \\
\label{eq:varchi}
\delta\chi=\:&-\frac{1}{2 \sqrt{2}}\slashed F \epsilon +\OO(\a^2)\:,
\end{align}
\eseq
where $\epsilon$  and $\chi$ are Majorana-Weyl spinors of positive chirality, $H_M = \frac{1}{2} H_{MNP} \Gamma^{NP}$ and
$ \slashed A_p = \frac{1}{p!} A_{M_1 \ldots M_p} \Gamma^{M_1 \ldots M_p}$.
We shall be using  connections with torsion defined as $\nabla_M^{\pm} = \nabla_M \pm \frac{1}{4} H_{M} $.

It is also  convenient  to introduce the modified dilatino variation 
\begin{equation}
\label{eq:varmoddil}
\delta\rho = \Gamma^M (\delta \psi_M)  - \delta \lambda  =  \slashed D \epsilon  
=\left(\slashed\nabla-\slashed\p\phi+\frac{1}{4}\slashed H-\frac{\a}{8}\slashed\Sigma\right)\epsilon+\OO(\a^2) \, . 
\end{equation}
With non-trivial gaugino bilinears the bosonic action \eqref{eq:Lagr0} and the bosonic equations of motion \eqref{eq:eom1}-\eqref{eq:eom4}  have
the same form with the replacement  \cite{Bergshoeff1989439, LopesCardoso:2003sp}
\begin{equation}
\label{Hshift}
H\rightarrow\bar H =H+\frac{\a}{2}\Sigma\:.
\end{equation}
Notice that, on the contrary, the presence of a gaugino condensate does not affect the Bianchi identity at $\OO(\a)$
 \begin{equation}
 \label{BIeq}
\d H - \frac{\a}{4} \left( \tr (R^-)^2 - \tr F^2 \right) = 0 \, .
\end{equation}

We will mostly be interested in supersymmetric solutions and, as usual, we will determine them by solving the supersymmetry 
conditions  \eqref{eq:varpsi}-\eqref{eq:varchi}. For zero-condensate  a solution of the supersymmetry constraints that  satisfies
the Bianchi identities for  $H$ is also a solution of the whole set of equations of motion \cite{Gauntlett:2002sc}. In presence of a condensate we do not
expect this to be true, and we would like to derive the set of extra constraints  required for a solution of the supersymmetry variations  to be also a  solution of the equations of motion. 
We will address the question by two different methods.  One  approach consists in extending 
the standard integrability
arguments of  \cite{Gauntlett:2002sc, Lust:2008zd}. 
The other  is  based on a generalisation of the Lichnerowicz theorem as in \cite{Coimbra:2014qaa}. The idea is to 
rewrite the ten-dimensional action as the sum of a term that is the square of the supersymmetry variations and some 
extra terms involving $\Sigma$.  
Even if this approach is less general than the integrability conditions, due to the use of integration by parts, it could provide an interesting way of computing solutions.

\subsection{Integrability of the supersymmetry equations}
\label{sec:int}
 
In this section we provide a general analysis of  the relation between supersymmetry and equations of motion in terms of the integrability conditions of the supersymmetry variations. Strictly speaking, gaugino condensates only make sense in the lower dimensional effective actions obtained from compactification. However,  we perform the analysis of integrability in the full ten-dimensional theory, treating the condensate as a formal object.  The aim is  to derive the most general set of constraints, which can then be applied to specific compactifications.

In absence of a condensate, it is possible to build combinations of squares of the supersymmetry variations that reproduce the equations of motion \cite{Gauntlett:2002sc, Lust:2008zd}

\begin{eqnarray}
\label{eq:BPS2}
& &  \Gamma^MD^0_{[N}D^0_{M]}\epsilon-\frac{1}{2}D^0_N(\mathcal{P}^0\epsilon)+\frac{1}{2}\mathcal{P}^0D^0_N\epsilon 
 =   -\frac{1}{4}\mathcal{E}^0_{NP}  \Gamma^P\epsilon  +\frac{1}{8}\mathcal{B}^0_{NP}  \Gamma^P\epsilon   + \frac{1}{2}  \iota_N  \d H^0  \epsilon \, ,  \nonumber  \\
&& \slashed D^0 \slashed D^0 \epsilon - ( D^{0 \, M}  - 2 \partial^M \phi) D^0_M \epsilon  = -\frac{1}{8} {\mathcal D}^0 \epsilon + \frac{1}{4} \d H^0 \epsilon  \, , 
\end{eqnarray}
where $\mathcal{E}^0_{NP}$,  $\mathcal{B}^0_{NP}$ and ${\mathcal D}^0$ denote the Einstein, $B$-field and dilaton equations of motion \eqref{eq:eom1} - \eqref{eq:eom3}.  $\d H$ is the Bianchi identity.\footnote{Note that, for simplicity,   we have ignored the contribution of the gauge fields 
in \eqref{eq:BPS2}.  As the gauge sector does not see the condensate at this order in $\a$, we are free to ignore it  in this computation.}
Since the left-hand side of both  equations in  \eqref{eq:BPS2} vanishes on supersymmetric solutions, 
the equations of motion are also satisfied, provided the Bianchi identity holds.

When the condensate is included, the first equation in \eqref{eq:BPS2} becomes
\begin{align}
&  \Gamma^MD_{[N}D_{M]}\epsilon-\frac{1}{2}D_N(\mathcal{P} \epsilon)+\frac{1}{2}\mathcal{P} D_N\epsilon 
 =   -\frac{1}{4}\mathcal{E}_{NP}  \Gamma^P\epsilon  +\frac{1}{8}\mathcal{B}_{NP}  \Gamma^P\epsilon  \nonumber \\
& \qquad \qquad   + \frac{1}{2}  \iota_N  \d H  \epsilon -  \frac{\a}{32} A_N(\Sigma) \epsilon  \, ,  
\label{eq:Int1A}
\end{align}
where now $\mathcal{E}_{NM}$,  $\mathcal{B}_{NM}$, ${\mathcal D}$ are the Einstein, $B$-field and dilaton equations of motion with non-zero condensate, and
the extra term $A_N(\Sigma)$ is given by 
\begin{align}
A_N(\Sigma) \epsilon &= 
A_{NP}\Gamma^P \epsilon +A_{NPQR}\Gamma^{PQR}\epsilon+ A_{NPQRST}\Gamma^{PQRST}\epsilon \nonumber\\
&=  \big[ e^{2 \phi} \nabla^M (e^{-2 \phi} \Sigma_{MNP} )+ \frac{1}{2} \Sigma_{NRS} {H^{RS}}_{P} + H \Sigma \delta_{NP} \big] \Gamma^P \epsilon  \nonumber \\
& \quad  +\frac{1}{2} \big[   e^{2 \phi} \nabla^M (e^{-2 \phi} \Sigma_{M PQ} ) \delta_{N R} + \frac{1}{3}  e^{ {-} 2 \phi} \nabla_N  (e^{ 2 \phi} \Sigma_{P QR} )   \nonumber \\
& \, \qquad +   \nabla_R  \Sigma_{NPQ}  -  H_{P ST} {\Sigma_Q}^{ST}  \delta_{NR} \big] \Gamma^{PQR} \epsilon  \nonumber \\
& \quad  - \frac{1}{6} \big[ \nabla_S \Sigma_{PQR} \delta_{NT}  + \frac{1}{2} H_{NPQ} \Sigma_{RST} \big] \Gamma^{PQRST} \epsilon \, . 
\label{eq:An0}
\end{align}  
Note that the tensors $\{A_{NM},A_{NPQR},A_{NPQRST}\}$  do not have any symmetry property. 
The analogue of the second equation in \eqref{eq:BPS2}  contains  extra terms in $\Sigma$ 
that  cancel non-tensorial terms
\begin{align}
\hspace{-0.3cm}& \slashed D  \slashed D \epsilon - ( D^{0 \, M}  - 2 \partial^M \phi  - \frac{\a}{16} \slashed \Sigma \Gamma^M - \frac{\a}{4} \Sigma^M ) D_M \epsilon  = -\frac{1}{8} {\mathcal D} \epsilon + \frac{1}{4} \d H \epsilon -  \frac{\a}{32}   B(\Sigma) \epsilon  \, , 
\label{eq:Int2A}
\end{align}
where the extra contribution from $\Sigma$ is 
\begin{align}
B (\Sigma) \epsilon & = B\epsilon+B_{NP}\Gamma^{NP} \epsilon+B_{MNPQ}\Gamma^{MNPQ}\epsilon \nonumber \\
 &=  6 H \lrcorner \Sigma \epsilon + 3 \big[  e^{2 \phi} \nabla^M (e^{-2 \phi} \Sigma_{MNP} )+ H_{NQR}  {\Sigma_P}^{QR}  \big] \Gamma^{NP} \epsilon  \nonumber \\
& \quad   + \frac{1}{3}\big[e^{-2\phi} \nabla_M\left(e^{2\phi} \Sigma_{NPQ}\right) -\frac{3}{2} \,H_{MNS} {\Sigma_{PQ}}^{S}\big]  \Gamma^{MNPQ }
\epsilon \, . 
\label{eq:B}
\end{align}
The left-hand sides of  \eqref{eq:Int1A} and \eqref{eq:Int2A} still vanish because of the supersymmetry variations, but now the analysis of the right-hand sides is more
involved.   For zero condensate, after imposing the Bianchi identity,  the only terms left are 
$\mathcal{E}^0_{NM}$ and  $\mathcal{B}^0_{NM}$  multiplying one gamma matrix, and they  must vanish separately because of their symmetry properties. 
In presence of condensate,  the extra terms $A(\Sigma)$ and $B(\Sigma)$ in   \eqref{eq:Int1A} and \eqref{eq:Int2A}
contain several terms involving different numbers of gamma matrices, which, in ten dimensions,  are not independent and hence cannot be set to
zero separately. 

To determine the set of independent constraints implied by the equations above we need to project them on a basis of spinors in ten dimensions. 
We follow closely the  discussion in  \cite{Tomasiello:2011eb}.
The supersymmetry parameter $\epsilon$ defines a  vector
\begin{equation}
\label{nulvec}
K^M =  \frac{1}{32} \bar{\epsilon} \Gamma^M \epsilon   \, 
\end{equation} 
that  is null and  annihilates the spinor $\epsilon$
\begin{equation}
\label{annK}
\slashed K  \epsilon = K^M \Gamma_M \epsilon  = 0  \, . 
\end{equation} 
Since $K$ is null, there are eight vectors orthogonal to it. We can  use $K$ to define a natural frame in $\mathbb{R}^{1,9}$.
This is given by $\hat{e}_- = K$,  $\hat{e}_\alpha$ , with $\alpha = 1, \ldots, 8$,  spanning the eight directions orthogonal to $K$, 
and another vector $e_+ = \tilde K$ that is not orthogonal to $K$
\begin{equation}
\hat{e}_- \cdot \hat{e}_+ = \frac{1}{2} \, , \qquad  \hat{e}_\pm \cdot \hat{e}_\pm = 0 \, , \qquad \hat{e}_\pm  \cdot \hat{e}_\alpha = 0  \, . 
\end{equation} 
 The ten-dimensional gamma matrices can be taken to be real\footnote{This means that $\Gamma_M^T = \Gamma^0  \Gamma_M \Gamma^0$.} 
 and decomposed as 
\begin{equation} 
\Gamma^{\pm} =\gamma_{(2)}^\pm\otimes \mathbb{I}  \qquad 
\Gamma^{\alpha} = \tilde\gamma_{(2)}\otimes \hat \gamma^\alpha \, , 
\end{equation} 
where $\hat \gamma^\alpha$ are eight-dimensional gamma matrices and
\begin{equation}
\gamma_{(2)}^+  =2 \left( \begin{array}{cc} 0 & 0 \\ 1 & 0 \end{array} \right)\:,\;\;\;\gamma_{(2)}^-= 2 \left( \begin{array}{cc} 0 & 1 \\ 0 & 0 \end{array} \right) \, . 
\end{equation}
The  two-dimensional chirality operator is 
$\tilde\gamma_{(2)} = \frac{1}{4}  (\gamma_{(2)}^-\gamma_{(2)}^+ -\gamma_{(2)}^+\gamma_{(2)}^- ) = \sigma_3$ 
and  $\hat \gamma_{(8)} =  \prod_{\alpha=1}^8 \hat \gamma^\alpha$  is  the eight-dimensional one.  The
  ten-dimensional  chirality is then  $\Gamma_{(10)} = \Gamma^0 \ldots \Gamma^9 = \tilde \gamma_{(2)} \otimes \hat \gamma_{(8)}$. 
In this basis,   \eqref{annK} becomes
\begin{equation}
\Gamma^+ \epsilon = \Gamma_- \epsilon = 0 \, ,
\end{equation}
which implies that the supersymmetry parameter decomposes as
\begin{equation}
\epsilon=\left( \begin{array}{c}0 \\ 1 \end{array} \right) \otimes   \eta =  | \downarrow >  \otimes  \, \eta \:,
\end{equation}
where $\eta$ is an eight-dimensional Majorana-Weil spinor of negative chirality, defining a $Spin(7)$ structure
in eight-dimensions (see appendix \ref{app:conv} for the definition of the relevant $G$-structures).  A basis for the 16-dimensional  Majorana-Weyl  spinors  is given by
\begin{equation}
\label{spinorbasis}
\epsilon \, ,
\qquad  
\omega_{\bf 7}^{\alpha \beta} \Gamma_{\alpha \beta} \epsilon =  | \downarrow >  \otimes \Pi^{\alpha \beta}_{{\bf 7} \, \, \gamma \delta}  \hat \gamma^{\gamma \delta} \eta \, , \qquad   \Gamma^{\alpha -} \epsilon  = 2 | \uparrow >  \otimes \hat  \gamma^\alpha \eta \, , 
 \end{equation}
where $ \Pi^{\alpha \beta}_{{\bf 7}  \, \, \gamma \delta}$ is the projector onto the representation ${\bf 7}$ of $Spin(7)$ and is given in  Appendix \ref{app:conv}. 
The first two terms above span the
space of the negative chirality spinors and the last term that of positive chirality ones.

The spinor $\epsilon$ defines a $Spin(7)  \ltimes \mathbb{R}^8$ structure. As usual the structure is equivalently given in terms of forms
that are bilinears in $\epsilon$. In this case  these are  
\beq
\label{10d structure} 
K  \qquad \qquad   \Psi  =  K \wedge \Phi_4 
\eeq
where $\Phi_4$ is the four-form associated to the $Spin(7)$ structure in the eight-dimensional space spanned by the vectors orthogonal to $K$.

In order to find the set of independent integrability conditions, we have to decompose  \eqref{eq:Int1A} and \eqref{eq:Int2A}  on the ten-dimensional basis 
\eqref{spinorbasis}. 
  We always assume that the Bianchi identity is satisfied
and that the left-hand sides vanish because of supersymmetry. 
We consider first the dilatino equation, \eqref{eq:Int2A}.  Clearly only the term of  $B(\Sigma)$  proportional to the singlet  can 
contribute to  the dilaton equation of motion.  Then, plugging \eqref{spinorbasis} into  \eqref{eq:Int2A}, we obtain that 
for the dilaton equation of motion to be satisfied  it is sufficient to require that 
\begin{align}
\label{eq:intA}
B+4\,B_{+-}+B_{\alpha \beta \gamma \delta }\Phi^{\alpha \beta \gamma \delta}=0 \, ,
\end{align}
where the tensors  $B$, $B_{+-}$ and $B_{\alpha \beta \gamma \delta }$ are defined in \eqref{eq:B}.  
 
Next let us consider  the Einstein and $B$-field equations. 
Decomposing \eqref{eq:Int1A}  in the spinorial basis \eqref{spinorbasis}, 
we find that the components $\mathcal{E}_{N-}$ and $\mathcal{B}_{N-}$ vanish if   
\begin{equation}
\label{eq:intB}
A_{N-}+5\,A_{N[\alpha\beta\gamma\kappa-]}\Phi^{\alpha\beta\gamma\kappa}=0 \, , 
\end{equation}
and, similarly, the components  $\mathcal{E}_{N \alpha}$ and $\mathcal{B}_{N \alpha}$ are implied if 
\begin{align}
\label{eq:An}
A_{N\alpha}+12\,A_{N[\alpha+-]}-\,A_{N \beta\gamma \delta }{\Phi^{\beta\gamma \delta}}_\alpha 
   +5\,A_{N[\beta\gamma\delta \epsilon \alpha]}\Phi^{\beta\gamma\delta \epsilon}-40\,A_{N[+ - \beta\gamma \delta ]}{\Phi^{\beta\gamma \delta}}_\alpha=0 \, . 
\end{align}
Notice that, since $\Gamma^+ \epsilon =0$,  the $+ +$ component of  the Einstein equation is always projected out in \eqref{eq:Int1A} \cite{Giusto:2013rxa}.
Imposing these constraints on $\Sigma$, we find again that supersymmetry and the Bianchi identity for $H$ imply all other equations of motion with the exclusion of
\begin{align}
\mathcal{E}_{++} = 0  \, . 
\end{align}

Note that the conditions  \eqref{eq:intA} and \eqref{eq:intB}   can be written in a base
independent way using the structure  $K$,  $\tilde K$  and $\Psi$ 
\bseq
\begin{align}
\label{intcB}
&B+4\, \tilde{K}^M  K^N B_{MN }+ 2  B_{NPQR } \tilde{K}_M\Phi^{MNPQR}=0   \\
\label{intcAp}
& A_{NP} K^P + A_{NPQRTS}\Psi^{PQRTS}=0 \, .
\end{align}
\eseq

\subsection{Supersymmetry and equations of motion}
\label{sec:EOM1}

In this section we describe an alternative approach to study the relation between supersymmetry variations and  equations of motion  
that  generalisese  Lichnerowicz theorem. 

Let us consider first the case when the condensate is zero.
The starting point is the Bismut-Lichnerowicz identity \cite{Coimbra:2014qaa}. Provided,  the Bianchi identity  is satisfied, the Bismut-Lichnerowicz identity allows to write the bosonic Lagrangian \eqref{eq:Lagr0} as 
\begin{equation}
\label{eq:nocondBismut}
\frac{1}{4}\mathcal{L}_b\epsilon+\OO(\a^2) = D^0_M D^{0\, M} \epsilon  - \slashed D^0 \slashed D^0 \epsilon+ \frac{\a}{16}\left(\tr\,\slashed F\slashed F\epsilon-\tr\,\slashed R^-\slashed R^-\epsilon\right)- 2\nabla^M\phi D^0_M\epsilon \, , 
\end{equation}
where $D^0_M$ and $\slashed D^0$  are the gravitino and modified dilatino equations, \eqref{eq:varpsi} and \eqref{eq:varmoddil}, with zero 
condensate. $R^-$ is the curvature two-form derived from the torsionful connection $\nabla^-$ (note that the gravitino variation involves $\nabla^+$). Multiplying \eqref{eq:nocondBismut} by  $e^{-2\phi}\epsilon^\dagger$ and integrating  it, gives the action
\begin{align} 
S_b&=-4\,\int_{M_{10}}\sqrt{-g}e^{-2\phi}\Big[\epsilon^\dagger\slashed D^0 \slashed D^0 \epsilon-\epsilon^\dagger D^0_M D^{0\, M} \epsilon\notag\\
&+2\nabla^M\phi\,\epsilon^\dagger D^0_M\epsilon-\frac{\a}{16}\left(\tr\,\epsilon^\dagger\slashed F\slashed F\epsilon-\tr\,\epsilon^\dagger\slashed R^-\slashed R^-\epsilon\right)\Big]+\OO(\a^2)\:,
\label{eq:LichAction}
\end{align}
where we assumed that $\epsilon^\dagger\epsilon=1$. 

We would like  to write the action as a BPS-squared expression, since its  variations will give  the equations of motion  in terms of the
supersymmetry variations. 
If the theory were Euclidean, we could integrate \eqref{eq:LichAction} by parts, and end up with such an action.  Unfortunately,  when the metric has  Lorentzian signature   it  is not possible, in general,   to reconstruct the supersymmetry operators $D^0_M$ and $\slashed D^0$ after the integration by parts.
Since the problematic terms always involve components of the fields with one leg along the time direction,  one can restrict to solutions where none of the fields
has components of this type. This means
\begin{equation}
\label{eq:staticFields}
H_{0MN}=\Sigma_{0MN}=F_{0M}=0\, ,
\end{equation}
and, for the metric\footnote{Note that given  \eqref{eq:staticFields} and \eqref{eq:staticMetric} for $H$ and the metric, $R^-$ has no temporal legs.}, 
\begin{equation}
\label{eq:staticMetric}
\d s_{10}^2=-e^{2A}\d t^2+g_{mp}\d x^m\d x^p\:,
\end{equation}
where $A=A(x^m)$, $g=g(x^m)$ and $\{x^m,x^n,..\}$ denote spatial coordinates. 
Note also that, in order to perform the integration by parts, we assume that the fields vanish at infinity where the metric is flat.\footnote{Without this assumption there might be unwanted boundary terms. The assumption holds for Minkowski compactifications but  excludes certain types of solutions, like domain walls.}
Under these assumptions  the action can be integrated by parts to give 
\begin{align}
\label{eq:nocondBismut2}
\int_{M_{10}}e^{-2\phi}\mathcal{L}_b = & 4  \int_{M_{10}}e^{-2\phi} [(\slashed D^0\epsilon)^\dagger\slashed D^0\epsilon-(D^0_M\epsilon)^\dagger D^{0 \, M}\epsilon \nonumber \\ 
& +\frac{\a}{16}\left(\tr\,\epsilon^\dagger\slashed F\slashed F\epsilon-\tr\,\epsilon^\dagger\slashed R^-\slashed R^-\epsilon\right)] +\OO(\a^2)\:.
\end{align}
The variations of  \eqref{eq:nocondBismut2} with respect to the metric, dilaton and $B$-field give the corresponding equations of motion written in terms
of the supersymmetry conditions. Thus  solutions of the supersymmetry constraints also automatically  solve the equations of motion.\footnote{
The variation of the second line in \eqref{eq:nocondBismut2} also vanishes. Its variation is proportional to 
\begin{equation*}
\slashed R^-_{MN}\,\epsilon=R^-_{PQMN}\Gamma^{PQ}\epsilon=R^+_{MNPQ}\Gamma^{PQ}\epsilon+\OO(\a)=\OO(\a)\:, 
\end{equation*}
which vanishes on supersymmetric solutions  to the appropriate order in $\a$. }

For non-trivial  condensate the same analysis gives, up to  $\OO(\a^2)$ terms, 
\bseq
\begin{align}
S_b&=BPS^2+\frac{\a}{8}\int_{M_{10}}\sqrt{-g}e^{-2\phi}\Big[(\slashed D\epsilon)^\dagger\slashed\Sigma\epsilon+(\slashed\Sigma\epsilon)^\dagger\slashed D\epsilon\notag\\
& \quad  +\frac{1}{2}\left((D_M\epsilon)^\dagger\slashed\Sigma\Gamma^M\epsilon+(\slashed\Sigma\Gamma^M\epsilon)^\dagger D_M\epsilon\right)-H\lrcorner\Sigma\Big]\:,
\label{eq:actionCond}
\end{align}
where $H\lrcorner\Sigma=\frac{1}{3!}H^{MNP}\Sigma_{MNP}$, and  $BPS^2$  denotes the part of the action that can be written as the square of
the supersymmetry variations 
\begin{equation}
\label{bpsterm}
BPS^2=\int_{M_{10}}\sqrt{-g}e^{-2\phi} \Big[ (\slashed D\epsilon)^\dagger\slashed D\epsilon-(D_M\epsilon)^\dagger D^M\epsilon+\frac{\a}{16}\left(\tr\,\epsilon^\dagger\slashed F\slashed F\epsilon-\tr\,\epsilon^\dagger\slashed R^-\slashed R^-\epsilon\right) \Big] \,.
\end{equation}
\eseq
Note that because of the extra terms involving $\Sigma$ on the right-hand side of \eqref{eq:actionCond},  the action can no longer be written as a BPS-squared expression.  The equations of motion for $g_{MN}$, $\phi$ and $B_{MN}$ obtained 
 by varying  \eqref{eq:actionCond}  will contain a term coming from its  $BPS$ part, which vanishes for
supersymmetric configurations, and terms in $\Sigma$,  which will provide additional  constraints to be imposed on the supersymmetric solution.
These extra terms have a  relatively simple structure  and in particular do not contain the curvatures $F$ and $R^-$. More concretely, the extra terms in the
dilaton and $B$-field equations of motion are respectively 
\bseq
\begin{align}
\label{eq:dilatonEOM}
& H_{MNP}\Sigma^{MNP}+e^{2\phi}\nabla^M\left(e^{-2\phi}\epsilon^\dagger\Gamma_{MNPQ}\Sigma^{NPQ}\epsilon\right)=0+\OO(\a)\, , \\ 
 \label{eq:fluxEOM}
& e^{2\phi}\nabla^M\left(e^{-2\phi}\,\epsilon^\dagger\{\Gamma_{MPQ},\slashed\Sigma\}\epsilon\right)=0+\OO(\a)\:.
\end{align}
\eseq
Both equations are only linear in  derivatives.
Note the resemblance between \eqref{eq:fluxEOM}  and the usual flux equation of motion \eqref{eq:eom3}.
The  metric equation of motion evaluated on a supersymmetric solutions 
gives the  equation\footnote{We also need  to vary the vielbeine in the gamma matrices as   $\delta\Gamma^M={\delta e_A}^M\,\Gamma^A$ 
and in the gaugino bilinear 
$\Sigma_{MNP}= \bar\chi\Gamma_{MNP}\chi = \bar\chi\Gamma_{ABC}\chi  {e_M}^A{e_N}^B{e_P}^C$.} 
\begin{align}
&e^{2\phi}\nabla_N\left[e^{-2\phi}\left(\epsilon^\dagger\Gamma^{N(P}\slashed\Sigma\Gamma^{M)}\epsilon-2g^{MP}\epsilon^\dagger\Gamma^N\slashed\Sigma\epsilon+2g^{N(M}\epsilon^\dagger\Gamma^{P)}\slashed\Sigma\epsilon\right)\right]=\nabla^{(M}\phi\,\epsilon^\dagger\Gamma^{P)}\slashed\Sigma\epsilon\notag\\
&+2\,g^{MP}H\lrcorner\Sigma-\,{H_{RS}}^{(M}\Sigma^{P)RS}-3\,{H_{NR}}^{(M}\epsilon^\dagger\Gamma^{P)N}\slashed\Sigma\Gamma^R\epsilon+\OO(\a)\:,
\label{eq:ModEinstein}
\end{align}

The equations above can  can be further simplified, using the fact that 
\begin{equation}
\label{eq:bcH}
H=\d\phi=0+\OO(\a)\:.
\end{equation}
This follows from imposing that the fields must 
vanish at infinity. 
 Indeed, substracting $1/4$ of the trace of \eqref{eq:eom1} from \eqref{eq:eom2}  we find 
\bseq
\begin{equation}
\label{eq:moddileq}
\frac{1}{2}\nabla^2\phi-(\nabla\phi)^2+\frac{1}{4}\,H^2=0+\OO(\a)\:,
\end{equation}
or, alternatively, 
\begin{equation}
e^{2\phi}\nabla^M\left(e^{-2\phi}\nabla_M\phi\right)+\frac{1}{2}\,H^2=0+\OO(\a)\:.
\end{equation}
\eseq
Multiplying by $e^{-2\phi}$ and integrating over spacetime then gives
\begin{equation}
\int_{M_{10}} e^{-2\phi} H^2=0+\OO(\a)\:.
\end{equation}
 As the integrand is positive, it has to vanish point-wise, which implies  \eqref{eq:bcH}.
 Integrating \eqref{eq:moddileq}, we also find that $\d\phi=0+\OO(\a)$. 
Using this result, the equations  \eqref{eq:dilatonEOM} -\eqref{eq:ModEinstein}  become (modulo $\OO(\a)$ terms)
\bseq
\begin{align}
\label{simpeq1}
& \tilde{\Psi}^{MNPQ}\,\nabla_{[M}\Sigma_{NPQ]}=0 \, , \\
\label{simpeq2}
& \nabla^P\Sigma_{PMN} + \,  \tilde{\Psi}_{PQ R  [M} \nabla^P {\Sigma^{QR}}_{ N]} - \frac{1}{2}  \tilde{\Psi}_{MNPQ} \nabla^R {\Sigma_R}^{PQ} =0 \, , \\
\label{simpeq3}
&  \tilde{\Psi}_{MPQ(R}\nabla^M{\Sigma_{N)}}^{PQ}-\frac{1}{6}{ \tilde{\Psi}^{TPQ}}_{(R}\nabla_{N)}\Sigma_{TPQ}=0\:,
\end{align}
\eseq
 where, for simplicity of notation, we have defined $\tilde{\Psi}_{MNPQ}=\epsilon^\dagger\Gamma_{MNPQ}\epsilon$.

As in the previous section, to extract the set of independent constraints from \eqref{simpeq1}-\eqref{simpeq3} we need to decompose them on
the spinorial basis \eqref{spinorbasis}.
It is straightforward to see that the only non-zero component of $\tilde{\Psi}$ is purely in the eight  directions orthogonal to the vector  $K$ 
and reduces to the $Spin(7)$ invariant form
\begin{equation}
\tilde{\Psi}_{\pm MNP}=0\:,\;\;\;  \tilde{\Psi}_{\alpha \beta \gamma \delta} =  \Phi_{\alpha \beta \gamma \delta} \, . 
\end{equation}
The condition that  there are no components in the time direction, 
\eqref{eq:staticFields},  implies 
\begin{equation}
\label{sigmacompstat}
\Sigma_{+ \alpha \beta} =  \Sigma_{- \alpha \beta}\:, \qquad \Sigma_{+ - \alpha} = 0 \:.
\end{equation}
It is immediate to see that  \eqref{simpeq1} reduces to 
\begin{equation}
\label{eq:dilatonEOM2}
\Phi \lrcorner \d   \Sigma = 0 \, .
\end{equation}
Combining the $+ \alpha$ components of \eqref{simpeq2} and \eqref{simpeq3}, one finds
\bseq
\begin{eqnarray}
\label{eq:EOM1}
&&   \nabla^{\beta} \Sigma_{ \beta \alpha  +}  + \frac{1}{2}  \Phi_{\beta \gamma \delta \alpha} \nabla^\beta {\Sigma^{\gamma \delta}}_+  = 0 \, ,   \\ 
\label{eq:EOM2}
&&  \Phi_{\gamma\delta\epsilon\alpha}\nabla^\gamma{\Sigma^{\delta\epsilon}}_++\frac{1}{6}{\Phi_\alpha}^{\gamma\delta\epsilon}\nabla_+\Sigma_{\gamma\delta\epsilon} =0  \, . 
\end{eqnarray}
\eseq
Then the  $ \alpha, \beta$ components of \eqref{simpeq2} and \eqref{simpeq3} reduce to
\bseq
\begin{eqnarray}
\label{eq:EOM3}
&& \left(2 \nabla^+ \Sigma_{+\gamma\delta}+\nabla^\epsilon\Sigma_{\epsilon\gamma\delta}\right)\left(\delta^{\gamma\delta}_{\alpha\beta}-\frac{1}{2}{\Phi_{\alpha\beta}}^{\gamma\delta}\right) + \Phi_{\delta\epsilon\sigma[\alpha}\nabla^\delta{\Sigma_{\beta]}}^{\epsilon\sigma}= 0    \\
\label{eq:EOM4}
&& \Phi_{\gamma \delta \epsilon (\alpha} \nabla^\epsilon {\Sigma_{\beta)} }^{\gamma \delta} - \frac{1}{6} \Phi_{\gamma \delta \epsilon (\alpha} \nabla_{\beta)} \Sigma^{\gamma \delta \epsilon}  =0 \:.
\end{eqnarray}
\eseq

We would like to compare these conditions with those obtained from integrability in section \ref{sec:int}. 
Recall that  in our analysis we neglect terms  of order $\a^2$ in the equations of motion
and that   \eqref{eq:intA} -- \eqref{eq:An} are already of order $\a$.
Since for static solutions  $H=\d\phi=\OO(\a)$,  we can discard  all terms involving products of $H$ and $\Sigma$  
in  \eqref{eq:intA} -- \eqref{eq:An}. 
Using also \eqref{sigmacompstat} it is easy to see that  \eqref{eq:intA}  reduces to
\begin{equation}
\label{eq:dilaton}
\nabla_{\alpha}\Sigma_{\beta\gamma\sigma}\Phi^{\alpha\beta\gamma\sigma} =0 \, , 
\end{equation}  
which agrees with  \eqref{eq:dilatonEOM2}. 
For  \eqref{eq:intB} one can show that only the $N=\alpha$ component gives a non trivial condition
\begin{equation}
\label{eq:Int1}
\nabla^\beta\Sigma_{\beta\alpha-}-\frac{2}{3}{\Phi_\alpha}^{\beta\gamma\delta}\nabla_{[-}\Sigma_{\beta\gamma\delta]} =0\:.
\end{equation}
From \eqref{eq:An} we get two equations. One for the component $N=+$
\begin{equation}
\label{eq:Int2}
3\nabla^\beta\Sigma_{\beta\alpha+}+\frac{1}{2}\nabla_+\Sigma_{\beta\gamma\delta}{\Phi^{\beta\gamma\delta}}_\alpha-\frac{5}{2}\nabla_\beta\Sigma_{\gamma\delta+}{\Phi^{\beta\gamma\delta}}_\alpha=0\: , 
\end{equation}
and another for  the component $N=\beta$, whose symmetric and anti-symmetric parts are 
\bseq
\begin{align}
\label{eq:Int4}
\left(2\nabla^+\Sigma_{+\gamma\epsilon}+\nabla^\sigma\Sigma_{\sigma\gamma\epsilon}\right)\left(\delta_{\alpha\beta}^{\gamma\epsilon}-\frac{1}{2}{\Phi_{\alpha\beta}}^{\gamma\epsilon}\right)+\frac{1}{3}{\Phi^{\gamma\delta\epsilon}}_{[\alpha}\nabla_{\beta]}\Sigma_{\gamma\delta\epsilon} &=0\\
\label{eq:Int5}
{\Phi_{(\alpha}}^{\epsilon\gamma\delta}\nabla_{\vert\epsilon\vert}\Sigma_{\beta)\gamma\delta}&=0\:.
\end{align}
\eseq

Using \eqref{eq:Int1} to get rid of the first term in \eqref{eq:Int2}, we find precisely \eqref{eq:EOM2}. This can then be inserted back into \eqref{eq:Int1} to obtain  \eqref{eq:EOM1}.
To compare the remaining equations, we need to take into account that, in order to derive the action \eqref{eq:actionCond}, 
we assumed that  the ten-dimensional spinor $\epsilon$ has constant norm. This implies  that 
\begin{equation}
\label{Phicondst}
\Phi_{\alpha\beta\gamma\delta}\Sigma^{\beta\gamma\delta}=0
\end{equation}
on the supersymmetric locus, as it can be shown  from 
\begin{equation}
0=\p_M\left(\epsilon^\dagger\epsilon\right)=\left(\nabla_M^{H^+}\epsilon\right)^\dagger\epsilon+\epsilon^\dagger\nabla_M^{H^+}\epsilon=\frac{\a}{48}\Psi_{MNPQ}\Sigma^{NPQ}\:,
\end{equation}
where we have used \eqref{eq:varpsi} and $H^+=H+\frac{\alpha^\prime}{4}\Sigma$. 
Using \eqref{Phicondst},  we see that \eqref{eq:EOM4} and \eqref{eq:Int5} coincide, while \eqref{eq:Int4} reduces to 
\begin{equation}
\left(2\nabla^+\Sigma_{+\gamma\epsilon}+\nabla^\sigma\Sigma_{\sigma\gamma\epsilon}\right)\left(\delta_{\alpha\beta}^{\gamma\epsilon}-\frac{1}{2}{\Phi_{\alpha\beta}}^{\gamma\epsilon}\right)=0\:.
\end{equation}
This is to be compared with \eqref{eq:EOM3}.  It can be checked that the projection of the two-form $ \Phi_{\delta\epsilon\sigma[\alpha}\nabla^\delta{\Sigma_{\beta]}}^{\epsilon\sigma}$ into $\bf 21$ vanishes, and hence  $\bf 7$ is its only surviving $Spin(7)$ representation. It can further be shown that this is the same as the projection to $\bf 7$ of ${\Phi^{\gamma\delta\epsilon}}_{[\alpha}\nabla_{\beta]}\Sigma_{\gamma\delta\epsilon}$.\footnote{To show this, note that $\nabla \Sigma$ - an object with 1+3 indices corresponds to $\bf 8 \times (35 + 21) \rightarrow 7+ 21 + 35 + 35 + 21 + 35 + 105 + 189$. But then one contracts it with $\Phi$ leaving only two free indices. In other words every such  contraction is a projection into $\bf 35 + 21 + 7$ (there could also be a trace part, i.e. $\bf 1$, but it is not there in the original decomposition, and as mentioned is vanishing by a separate condition \eqref{eq:dilatonEOM2} and \eqref{eq:Int1}). Note that 2 different projections appear $\Phi_{\gamma \delta \epsilon \alpha} \nabla^\epsilon {\Sigma_{\beta} }^{\gamma \delta}$ and $ \Phi_{\gamma \delta \epsilon \alpha} \nabla_{\beta} \Sigma^{\gamma \delta \epsilon}=0$ at the supersymmetric locus. But there is only a single $\bf 7$ so it must be the same in both projections.} The latter vanishes by virtue of  \eqref{Phicondst} and $\nabla_M \epsilon \sim \mathcal{O} (\a)$.

To summarise, for static solutions vanishing at infinity,  both set of constraints  \eqref{eq:EOM3} - \eqref{eq:EOM4} and  \eqref{eq:Int4} - \eqref{eq:Int5}  reduce to
\bseq
\begin{align}
\label{eq:Int6}
\Pi_7\left(2\nabla^+\Sigma_{+\gamma\epsilon}+\nabla^\sigma\Sigma_{\sigma\gamma\epsilon}\right)  \equiv \left(2\nabla^+\Sigma_{+\gamma\epsilon}+\nabla^\sigma\Sigma_{\sigma\gamma\epsilon}\right)\left(\delta_{\alpha\beta}^{\gamma\epsilon}-\frac{1}{2}{\Phi_{\alpha\beta}}^{\gamma\epsilon}\right) &=0\\
\label{eq:Int7}
{\Phi_{(\alpha}}^{\epsilon\gamma\delta}\nabla_{\vert\epsilon\vert}\Sigma_{\beta)\gamma\delta} &=0\:.
\end{align}
\eseq
Note that only $(\d \Sigma)_{\bf 27}$ may still be non-vanishing.

Regardless of the above considerations, one may be tempted to impose by hand the 
stronger condition
\begin{equation}
\label{eq:closureSig}
\d\Sigma = 0 \, , 
\end{equation}
which can be interpreted  as a  ``Bianchi identity"  for the  condensate three-form. 
Indeed, we recall that when the condensate is turned on the  bosonic  action is obtained from the usual one by the replacement
\begin{equation}
\label{eq:repl}
H\rightarrow\bar{H} = H + \frac{\alpha^\prime}{2} \Sigma\:.
\end{equation}
With \eqref{eq:closureSig} imposed, $H$ and $\bar{H}$   satisfy the same Bianchi identity. As we shall see, this condition allows to construct solutions (possibly non-supersymmetric) with nontrivial condensate form a solution with $\Sigma = 0$.

\subsection{Solutions with Fake Supersymmetry}
\label{sec:SpecialSol1}

A similar approach to the integrability of  section \ref{sec:int}  provides a nice method to find solutions of the equations of motion that are not supersymmetric.  
Solutions of this type are in the spirit of \cite{Derendinger198565, Dine:1985rz, DERENDINGER1986365}, where it is shown that gaugino condensation provides a natural mechanism for supersymmetry breaking.

As already mentioned several times, the heterotic equations of motion with non trivial condensate are formally the same as for zero condensate, \eqref{eq:eom1} - \eqref{eq:eom4}, 
but with the replacement 
\begin{equation}
H\rightarrow\bar H =H+\frac{\a}{2}\Sigma\, .
\end{equation}

Let us  define "fake" supersymmetry variations  that have the same form as those for zero condensate \eqref{eq:varpsi}-\eqref{eq:varmoddil} with $H$  replace by $\bar{H}$
\bseq
\begin{align}
\label{eq:varpsifake}
\delta\psi_M =\:&     \bar{D}_M \epsilon = \nabla_M\epsilon+\frac{1}{4} \bar{H}_{M} \epsilon  +\OO(\a^2) \, , \\
\label{eq:vardilfake}
\delta\lambda =\:&  -\frac{\sqrt{2}}{4}\, \bar{\mathcal{P}} \epsilon = -\frac{\sqrt{2}}{4} \Big[\slashed\p\phi+\frac{1}{2} \bar{\slashed  H}  \Big] \epsilon+\OO(\a^2) \, , \\
\label{eq:varmoddilfake}
\delta\rho = \:&\bar{  \slashed D}  \epsilon   = \left(\slashed\nabla-\slashed\p\phi+\frac{1}{4} \bar{ \slashed H}  \right)\epsilon+\OO(\a^2) \, . 
\end{align}
\eseq

Then if we  square the equations above  as in  section \ref{sec:int} we find 
\bseq
\begin{align}
\label{eq:BPSbar2}
& \Gamma^M\bar D_{[N}\bar D_{M]}\epsilon-\frac{1}{2} \left[ \bar D_N ,  \mathcal{\bar P} \right] \epsilon 
 =   -\frac{1}{4}\mathcal{E}_{NP}  \Gamma^P\epsilon  +\frac{1}{8}\mathcal{B}_{NP}  \Gamma^P\epsilon   + \frac{1}{2}  \iota_N  \d  \bar H  \epsilon \, ,    \\
& \slashed{\bar D} \slashed{\bar D} \epsilon - (\bar D^{M}  - 2 \partial^M \phi)\bar D_M \epsilon  = -\frac{1}{8} {\mathcal D} \epsilon + \frac{1}{4} \d \bar H \epsilon   \, , 
\end{align}
\eseq
where  $\mathcal{E}_{NP}$,  $\mathcal{B}_{NP}$ and ${\mathcal D}$ are the Einstein, $B$-field and dilaton equations of motion 
with non-zero condensate.  Notice that the term involving the Bianchi identity contains a correction is $\Sigma$
\beq
\d \bar H = \d H + \frac{\a}{2} \d \Sigma \, .
\eeq
From \eqref{eq:BPSbar2} we see that a background satisfying
\beq
\label{fakesusy1}
\bar D_{M} \epsilon  = 0 \quad \quad   \slashed{\bar D} \epsilon = 0  \quad \quad  \mathcal{\bar P} \epsilon = 0 
\eeq
plus the  Bianchi identity $\d H=0$ and the closure of $\Sigma$ 
\beq
\label{fakesusy2}
\d \Sigma  =0 
\eeq
is also a solution of the equations of motion.  The advantage of this approach is that, to find non-supersymmetric solutions,  it is sufficient to solve susy-like first order equations, which
are generally simpler than the full equations of motion.  Generically the solutions will not be supersymmetric, hence the name fake supersymmetry. 

One can also ask what extra conditions must be imposed for the solution to be supersymmetric.  Comparing \eqref{eq:varpsifake} with  \eqref{eq:varpsi}  we see that the former reduces
to the latter if 
\begin{equation}
\label{eq:specialsol}
\Sigma_{NPQ}{\Gamma^{NPQ}}_M\epsilon=3\,\Sigma_{MNP}\Gamma^{NP}\epsilon \, . 
\end{equation}
This condition also implies that $\slashed \Sigma$ vanishes and hence \eqref{eq:vardilfake} and \eqref{eq:varmoddilfake}
also reduce to the supersymmetric ones.
The class of supersymmetric backgrounds obtained this way is more reduced than what we would obtain by solving the supersymmetry equations \eqref{eq:varpsi} - \eqref{eq:varmoddil}
plus the integrability conditions, but have the advantage that one only needs to solve first order equations.
We will return to some explicit solutions of this type in section \ref{sec:Ex}.

\section{Compactifications to four dimensions}
\label{sec:split46}

We consider first the  supersymmetry conditions for compactifications to four-dimen\-sional maximally symmetric spaces, Minkowski or Anti de Sitter.
The metric is of  the form   
\begin{equation}
\d s^2_{10}=e^{2\Delta}\d s^2_4+\d s^2_6 \, , 
\end{equation}
where the warp factor only depends on the internal coordinates and  the internal manifold  is  assumed to have 
$SU(3)$ structure. 

We denote four-dimensional indices by greek letters, $\mu, \nu, \ldots = 0, \ldots, 3$ and internal indices by latin ones, $m, n, \ldots =  1, \ldots, 6$.
The ten-dimensional gamma matrices decompose as
\begin{align}
\Gamma^\mu&= e^{\Delta} \gamma^\mu\otimes\bf 1\notag\\
\Gamma^m&=\gamma_{(4)} \otimes\gamma_m\:,
\end{align}
where $\gamma^\mu$  are four-dimensional gamma matrices of the unwarped metric and $\gamma_{(4)}$ is the chiral gamma in four dimensions. The six-dimensional matrices $\gamma^m$
are hermitian. The supersymmetry parameter decomposes as
\begin{equation}
\label{eq:spinordecomp}
\epsilon=e^A\zeta_+\otimes\eta_++e^A\zeta_-\otimes\eta_- \, ,
\end{equation}
where $\zeta_{\pm}$ and $\eta_\pm$ are Weyl spinors of positive and negative chirality in four and six dimensions, respectively
 (see appendix \ref{app:6d} for more conventions). 

The ten-dimensional gravitino splits as 
\begin{equation}
\chi_{(10)}=\chi^{(4)}_+\otimes\chi_++\chi^{(4)}_-\otimes\chi_-\:,
\end{equation}
where $\chi^{(4)}_\pm$ and $\chi_\pm$ are four- and six-dimensional Weyl spinors.  
In Calabi-Yau  compactifications,  the internal spinor $\chi$ is taken to be a singlet of the $SU(3)$ holonomy group in order to 
 have a massless four-dimensional gaugino. In this paper we are interested in the larger class of $SU(3)$ structure compactifications and we a-priori take $\chi_+$ to be a generic six-dimensional Weyl spinor.

Poincar\'e invariance in four dimensions forces the three-form flux $H$
and   the three-form condensate to have only non-trivial components in the internal space. We are still taking a ten-dimensional  expectation value,  $\<\Sigma_{MNP}\>$. We should also remember that the ten-dimensional gauginos are in the adjoint of $E_8 \times E_8$ or $SO(32)$. Denoting the external four-dimensional gauge group  as $G$, and the internal group as $H$ ($G$ is the stabilizer of $H$ in the ten-dimensional group), we may decompose the ten-dimensional product representation as  ${\bf 496} \otimes {\bf 496} \rightarrow \sum_i (R(G)_i, R(H)_i)$. Of course the details very much depend on the choice of $G$ and $H$, but in general the ten-dimensional trace over fermion bilinears will  break into a sum of many terms:
\begin{align}
\SSigma_{mnp}&=\langle {\rm tr} \bar\chi_{(10)}\Gamma_{mnp}\chi_{(10)}\rangle\notag\\
&=\sum_i \langle {\rm tr}_{R(G)_i} \bar\chi^{(4)}_+\chi^{(4)}_-\, \cdot {\rm tr}_{R(H)_i} \chi_+^\dagger\gamma_{mnp}\chi_- \rangle - \sum_i \langle {\rm tr}_{R(G)_i} \bar\chi^{(4)}_-\chi^{(4)}_+\cdot \, {\rm tr}_{R(H)_i} \chi_-^\dagger\gamma_{mnp}\chi_+ \rangle \notag\\
&=  - 2 \sum_i {\rm Re} \langle  \Lambda_i  \Sigma^i_{mnp} \rangle \, . 
\end{align}
Here we have defined internal three-forms   $\Sigma^i_{mnp}$  as
\begin{equation}
\label{eq:Sigma4}
\Sigma^i_{mnp}= {\rm tr}_{R(H)_i} \chi^\dagger_-\gamma_{mnp}\,\chi_+  \, , 
\end{equation}
and   a  four-dimensional condensate {\sl vector} $\Lambda_i$ as
\begin{equation}
\label{eq:Lambda4}
\Lambda_i =  {\rm tr}_{R(G)_i} \bar\chi^{(4)}_-\chi^{(4)}_+  \, . 
\end{equation}
From now on we shall suppress all the traces and the $G$ and $H$ representation indices. To simplify the notation, in the rest of this section we will  set $\langle\Sigma \rangle = \hat{\Sigma} = - 2  {\rm Re} (  \Lambda  \Sigma ) $.

As shown in appendix   \ref{app:deriv4d},  the ten-dimensional supersymmetry equations can be written as a set of  conditions on 
the forms $\Omega$ and $J$ defining the $SU(3)$ structure.  The set of independent equations are
\bseq
\begin{align}
\label{eq:4dBPS1}
H  \lrcorner \Omega = 2  i \bar{\mu} e^{- \Delta}   \qquad  i\frac{\a}{8} \hat{\Sigma} \lrcorner \Omega =  i  \bar{\mu}  e^{-\Delta} \:,
\end{align}
and  the differential conditions
\begin{align}
\label{eq:4dBPS4}
\d J&=\left(2\d\phi-4\d\Delta\right)\wedge J+*\bar H+3\,e^{-\Delta}{\rm Re}\left(\mu\,\Omega\right)\\
\label{eq:4dBPS5}
\d\Omega&=\left(2\d\phi-3\d\Delta\right)\wedge\Omega-ie^{-\Delta}\,\bar\mu\,J\wedge J \\
\label{eq:4dBPS3}
J\wedge\d J&=(\d\phi-\d\Delta)\wedge J\wedge J\\
\label{eq:4dBPS2}
\frac{\a}{4}*  \hat{\Sigma} \wedge J&=\d\Delta\wedge J\wedge J
\end{align}
\eseq
where we define $\bar H=H+\frac{\a}{2} \hat{\Sigma}$. A similar set of supersymmetry conditions was also derived in \cite{Held:2010az}. Note that the flux equation of motion is automatically satisfied
\begin{equation}
\d\left(e^{-2\phi+4\Delta}*\bar H\right)=0 \: ,
\end{equation}
while the integrability conditions  of section \ref{sec:int}  have to be imposed to ensure that also the  other equations of motion are satisfied. This generically gives non-trivial conditions on the three-form $\hat\Sigma$.

It is interesting to see whether the supersymmetry equations are enough to completely determine the solution. 
To this extent, we decompose $H$ and $\hat{\Sigma} $ in $SU(3)$ representations 
\bseq
\begin{align} 
H &= \frac{3}{2} {\rm Im}  (H_{\bf 1} \bar{\Omega}) +  H_{(3)} \wedge J + H_{\bf 6} \,  , \\
\hat{\Sigma}  &= \frac{3}{2} {\rm Im}  (\hat{\Sigma}_{\bf 1} \bar{\Omega}) +  \hat{\Sigma}_{(3)} \wedge J + \hat{\Sigma} _{\bf 6} \,  , 
\end{align}
\eseq
where $H_{(3)} = H_{\bf 3} + H_{\bf \bar{3}}$ and ($i$ is a complex index)
\bseq
\begin{align}
H_{\bf 1} &= \frac{1}{36}\Omega^{mnp}H_{mnp}\\
H_{(3)\, i}&=\frac{1}{4} H_{i m n} J^{mn} \\
H_{{\bf 6} \,mnp} &=H_{mnp}-\frac{3}{2}{\rm Im}(H_{\bf 1} \bar\Omega)_{mnp}-3H_{(3)m}J_{mp]}\:,
\end{align}
\eseq
and similarly for $\hat{\Sigma}$. Comparing the supersymmetry conditions \eqref{eq:4dBPS4}  - \eqref{eq:4dBPS2} with the torsion equations 
\bseq
\begin{align}
\label{dJstr}
\d J&=\frac{3}{2}{\rm Im}(W_{1} \bar\Omega)+W_4\wedge J+W_3\\
\label{dOmegastr}
\d\Omega &= W_1\,J\wedge J+W_2\wedge J+\bar W_5\wedge\Omega \, , 
\end{align}
\eseq
we can see how many of the $SU(3)$ irreducible representations are fixed by supersymmetry. We find the following table
\begin{center}
\begin{tabular}{c|c|c}
$SU(3)$ repr.  &  parameters &  susy eq. \\
\hline 
$\mathbf{1}$ & $W_{\mathbf{1}}$ \,   $e^{-\Delta}\mu$ \,  $\bar{H}_{\bf 1}$ \, $\hat{\Sigma}_{\bf 1} $ & 3 \\
$\mathbf{3+\bar 3}$ & $W_{4}$ \, $W_5$ \, $\d\phi $\, $\d\Delta$ \, $ \bar H_{\bf 3}$, $\bar H_{\bf \bar 3} $ \, $ \hat \Sigma_{\bf 3}$, $\hat \Sigma_{\bf \bar 3} $  & 4\\
$\mathbf{6}$ & $W_{3}$ \, $\bar H_{\bf 6}$ \, ${\hat{\Sigma}}_{\bf 6}$ & 1 \\
$\mathbf{8}$ & $W_2$ & 1
 \end{tabular} 
 \end{center}
 
From \eqref{dOmegastr} and \eqref{eq:4dBPS5} we immediately see that the $\bf 8$ representation in the torsion is set to zero
\beq
W_2 = 0 \, .
\eeq
It is also easy to check  that all the singlet representations are fixed 
\bseq
\begin{align}
& W_1  =  - i \bar\mu e^{- \Delta} \\
& \bar H_{\bf 1} = 3\,\bar\mu\,e^{- \Delta} \\
& \a\,{\rm Re} (\Lambda \Sigma)_{\bf 1} =   \frac{1}{3}\bar\mu\,e^{- \Delta} 
\end{align}
\eseq

On the other hand,  we are  left  with one undetermined  quantity in the $\mathbf{3+\bar 3}$ representation 
\bseq
\begin{align}
\label{3repr}
& W_4 =2 \d \phi - 4  \d \Delta \\ 
& W_5 = 2 \d \phi - 3 \d \Delta \\ 
& \bar{H}_\mathbf{3} = i(\d\phi - 3\,\d\Delta)^{(1,0)} \\
\label{3reprd}
& \alpha^\prime {\rm Re}( \Lambda \Sigma)_\mathbf{3} = 2 i (\d\Delta)^{(1,0)} \, 
\end{align}
\eseq
and also for the  $\mathbf{6}$, where   ${\rm Re}( \Lambda \Sigma)_{\bf 6}$ is left undetermined
\begin{align}
& W_3 = \d J_{\mathbf{6}}= (*\bar H)_{\mathbf{6}}\:.
\end{align}
To have a solution of  the equations of motion,  in addition to the above set of supersymmetry conditions, we must also satisfy 
the  conditions derived from the integrability and the Bianchi identity \eqref{BIeq}.  On the four-dimensional external space there is no $H$-flux and  the Bianchi identity is trivially zero. 
One might worry that whenever there is a non vanishing cosmological constant, the right-hand side of the Bianchi identity can give a non-trivial contribution in the four-dimensional spacetime, coming from the curvature of AdS. However such a piece is by \eqref{eq:4dBPS1} of $\OO(\a)$ and the corresponding contribution to the Bianchi identity is $\OO(\a^2)$. 
For the same reason (see   \eqref{3reprd}) there is no contribution from the warp factor at this order.
Then we are left with the purely six-dimensional  equation 
\begin{equation}
\d H=\frac{\a}{4}\left(\tr\,F\wedge F-\tr\, R^-\wedge R^-\right)\:.
\end{equation}

Up to now we did not make any assumption on the nature of  the internal six-dimensional space.
When the manifold is compact, which is the most relevant case  for string phenomenology,
one can show that  "fake supersymmetry"  solutions of the  type of section \ref{sec:SpecialSol1} are the only allowed ones.
Notice first that  \eqref{eq:4dBPS1} implies that  at zeroth order in $\a$ there are no $AdS_4$ solutions.\footnote{Compact solutions with condensates and non trivial $AdS_4$ parameter have been found before in the literature, see e.g. \cite{Held:2010az, Lechtenfeld:2010dr, Chatzistavrakidis:2012qb, Klaput:2012vv}. Note however that such solutions inevitably require $\OO(1/\a)$ effects or similar to make sense.} 
Then a  well-known no-go theorem  \cite{Gauntlett:2002sc} states that  at zeroth order in $\a$ warping, dilaton and $H$-flux vanish and 
 the internal geometry must be Calabi-Yau.  We denote the  CY  geometry  by  $(X_0,\Omega_0,J_0)$ where
\begin{equation}
\d\Omega_0=0\:,\;\;\;\d J_0=0\: ,
\end{equation}
and we  express the full solution as an expansion in $\a$ 
\bseq
\begin{align}
\Omega&=\Omega_0+\a\Omega_1+..\\
J&=J_0+\a J_1+..\:,
\end{align}
\eseq
and similarly for the other fields, where   $H$-flux, dilaton, warping and condensate have no  $(\a)^0$ terms. Consider then \eqref{eq:4dBPS5},  which,  at first order,  reads
\begin{equation}
\d\Omega_1=\left(2\d\phi_1-3\d\Delta_1\right)\wedge\Omega_0-i \bar\mu_1\,J_0\wedge J_0\:,
\end{equation}
If $\mu_1\neq0$, this equation implies  that $J_0\wedge J_0$ is an exact four-form, which is not possible on a compact Calabi-Yau three-fold of finite volume. We conclude that $\mu_1=0$ for  supersymmetric solutions, and we can drop the cosmological constant to the order we are working at. A similar result was also derived in \cite{Quigley:2015jia}.

Next we consider the integrability condition obtained from \eqref{eq:B}, which on the first order geometry reduces to
\begin{equation}
\label{eq:Int4dred1}
\nabla^{m}\p_m\Delta_1=0\:,
\end{equation}
where we have used the BPS equation \eqref{eq:4dBPS2}. It follows that $\Delta_1$ is constant on a compact geometry. The BPS conditions for such geometries can hence,  without loss of generality, be taken to be
\bseq
\begin{align}
\label{eq:MinkBPS1}
J\wedge\d J&=\d\phi\wedge J\wedge J\\
\label{eq:MinkBPS2}
\d J&=2\,\d\phi\wedge J+*\left(H+\frac{\a}{2}\hat\Sigma\right)\\
\label{eq:MinkBPS3}
\d\Omega&=2\,\d\phi\wedge\Omega\:,
\end{align}
\eseq
where $\hat\Sigma$ is some primitive $(2,1)+(1,2)$ form. Primitivity follows from $\d\Delta=0$ and \eqref{eq:4dBPS2}. Using the integrability conditions \eqref{eq:An0} and \eqref{eq:B} one can then show that $\hat \Sigma$ must satisfy 
\begin{equation}
\nabla_{[m}\hat\Sigma_{npq]}\gamma^{npq}\eta_+=0 \: , 
\end{equation}
which in turn implies   $\d \Sigma=0$. Hence we recover the fake supersymmetry conditions of section \eqref{sec:SpecialSol1}.

\section{Backgrounds with  three-dimensional Poincar\'e invariance}
\label{sec:split37}
In this section we consider compactifications to three dimensions.  The  ten-dimensional spacetime is a warped product  
of a maximally symmetric three-dimensional space $M_3$  and a seven-dimensional manifold $X_7$. The   metric is
\begin{equation}
\label{eq:compSeven}
\d s^2=e^{2\Delta}\d s_3^2+\d s_7^2\,, 
\end{equation}
where  the warp factor $\Delta$ only depends on  the coordinates on $X_7$.  The three-dimensional space can be either Minkowski or Anti de Sitter, while 
 $X_7$ is not necessarily compact and admits a $G_2$ structure. 
Taking a non compact  $X_7$ allows to describe  also 
 four-dimensional domain-wall solutions  and standard compactifications
to four dimensions. As we will see, generic 
 seven-dimensional compactifications allow for more components of fluxes and, more importantly for this paper, condensate configurations.

With the ansatz \eqref{eq:compSeven}, the ten-dimensional gamma matrices decompose as
\begin{align}
\Gamma_\mu&= e^{ \Delta}  \gamma_\mu \otimes\sigma_3 \otimes \mathbb{I}_{(8)} \notag\\
\Gamma_m&=  \mathbb{I}_{(2)}  \otimes \sigma_1 \otimes \gamma_m \:,
\label{3+7gammas}
\end{align}
where we  denote the three-dimensional and seven-dimensional indices by   $\{\mu, \nu,..\}$ and  $\{m,n,..\}$, respectively.
 $\gamma_\mu$  are chosen to be  real,  while  $\gamma_m$ are purely imaginary. 
The ten-dimensional chiral operator is  
 $ \Gamma_{(10)} = -  \mathbb{I}_{(2)}  \otimes\sigma_2 \otimes \mathbb{I}_{(8)} $.
 
 The ten-dimensional spinors   split accordingly. The supersymmetry parameter 
 $\epsilon$ decomposes as 
\begin{equation}
\label{eq:epsilon7split}
\epsilon= e^A \zeta \otimes\left( \begin{array}{c}1 \\  - i  \end{array} \right)\otimes\eta\: , 
\end{equation}
where $\zeta$ is a three-dimensional spinor and $\eta$ a seven-dimensional spinor  of unitary norm. 
The spinor  $\eta$ is globally defined and  defines a $G_2$ structure on $X_7$.

Similarly,  the gaugino takes the following form
\begin{equation}
\label{eq:chi7split} 
\chi_{(10)}= \frac{1}{\sqrt{2}} \xi  \otimes\left( \begin{array}{c}1 \\ -  i \end{array} \right)\otimes\chi \, ,
\end{equation}
where again $\xi$ and $\chi $ are a three- and seven-dimensional spinors.   The internal spinor $\chi$ may be expressed  in terms of the $G_2$ spinor $\eta$ as follows
\begin{equation}
\label{eq:chi7} 
\chi^{i} =c^{i} \eta + c_m^{i} \gamma_m \eta \, ,
\end{equation}
where $i$ is the internal gauge index, and $c^{i}$ and $c_m^{i}$ are generic real functions of the internal coordinates. The norm of  $\chi$ is 
$C= \chi^\dagger \chi = \tr (c^2  + c_m c_m)$.\footnote{As in section \ref{sec:split46}, all the traces and three- and seven-dimensional representations are suppressed. $C$ has to be understood as a vector. The same applies to $\Lambda$  the internal three-form  $\Sigma$ to be defined shortly.}

The split \eqref{eq:chi7split} allows for two non-vanishing components of the three-form $\Sigma_{MNP}$ that are  compatible with the three-dimensional 
Poincar\'e  symmetry,
one with  all indices  in the three-dimensional space and one with all indices in the internal space. 
Recall that  $\Sigma_{MNP}$   is a gaugino bilinear, and as in the previous section    $\<\Sigma_{MNP}\>$ has to be interpreted as ten-dimensional vev.
Using \eqref{3+7gammas}  and \eqref{eqapp:gammaprod} one can show that  in flat indices  
\begin{equation}
\label{Sigmaans}
\< \Sigma_{ABC} \>  =  \left\{ \begin{array}{lcl}
 \< \bar{\chi}_{(10)} \Gamma_{\alpha \beta \gamma} \chi_{(10)} \> =  \< C \, \Lambda \> \, \epsilon_{\alpha \beta \gamma}   \quad \alpha, \beta, \gamma =1,2,3 \\  
\< \bar{\chi}_{(10)}  \Gamma_{a b c} \chi_{(10)} \>   =  -  i   \< \Lambda  \Sigma_{a b c}  \> \, ,
\end{array} 
\right. 
\end{equation}
where we have defined $\Lambda =  \tr \bar \xi \xi$ and  $ \Sigma_{abc}  = \chi^\dagger \gamma_{abc} \chi$. The fact that the gaugino $\chi$ is Majorana ensures that  $\Sigma_{abc}$ is imaginary. 

Poincar\'e invariance also implies that the only non-trivial components of the $H$-flux are
\begin{equation}
\label{Hans}
H_{MNP} =  \left\{ \begin{array}{lcl}
H_{\mu \nu \rho} =  \tilde{H} \epsilon_{\mu \nu \rho}  \, , \\ 
H_{mnp}  \, . 
\end{array} 
\right. 
\end{equation}

The ansatz \eqref{eq:compSeven},  \eqref{Hans} and \eqref{Sigmaans} for the metric,  $H$-flux and condensate can be used, together with
the splitting \eqref{3+7gammas}  and \eqref{eq:epsilon7split}, to reduce the ten-dimensional supersymmetry variations to a set of equations
on the forms $\varphi$ and $\psi$ defining the $G_2$-structure on the internal manifold.\footnote{Definitions and conventions for the forms $\varphi$ and $\psi$ are given in appendix \ref{app:conv}. In appendix \ref{app:deriv}  we give the derivation of the equations below.} 
These consist of two algrebraic relations 
\bseq
\begin{align}
\label{eq:SUSY1}
& H\lrcorner\varphi =  e^{-3\Delta} (2 \mu \, e^{2 \Delta} - \frac{\a}{2} \Lambda C ) \, ,  \\
\label{eq:SUSY2}
& i \frac{\a}{4}\Lambda\Sigma\lrcorner\varphi  = e^{-3 \Delta} \, ( \tilde{H} + \frac{\a}{4} C \Lambda - 2 \mu e^{2 \Delta}) \, , 
\end{align}
\eseq
together with the differential conditions
\bseq
\begin{align}
\label{eq:SUSY5}
\d\varphi&=\left(2\d\phi-3\d\Delta\right)\wedge\varphi- ( \ast H-i\frac{\a}{2} \ast (\Lambda \Sigma)) + 2\mu  \,e^{-\Delta}  \,\psi \, \\
\label{eq:SUSY6}
\d\psi&=\left(2\d\phi-2\d\Delta\right)\wedge\psi  \, , \\
\label{eq:SUSY3}
\d \Delta & = i \frac{\a}{8}\Lambda\,\Sigma \lrcorner \psi  \, . 
\end{align}
\eseq
For consistency of the theory, we must also satisfy the Bianchi identity. The three-dimensional external part  is trivial. Since the warping  
by \eqref{eq:SUSY3} gives an $\OO(\a^2)$ contribution to the right-hand side,  the internal Bianchi identity is 
\begin{equation}
\d H=\frac{\a}{4}\left(\tr\,F\wedge F-\tr\, R^-\wedge R^-\right)\:.
\end{equation}

Decomposing $H$ (and similarly $\Sigma$)   in $G_2$ representations 
\bseq
\begin{align}
& H_{mnp} =  \frac{1}{7} H_{\bf 1} \phi_{mnp}   -\frac{1}{4} H_{\bf 7}^q  \psi_{qmnp}  + \frac{3}{2} H_{{\bf 27} \, q [m } \phi_{np]}^{\, \, \, \,  \, q}  \, ,   \\
& (\ast H)_{mnpq} =  \frac{1}{7}  H_{\bf 1} \psi_{mnpq}  +  H_{{\bf 7} [m}  \phi_{npq]}  - 2 H_{{\bf 27} \, e [r} \psi^{r}_{\, \,  npq]} \, ,
\end{align}
\eseq
with (and similarly for $\Sigma$) 
\beq
H_{\bf 1} =  \frac{1}{6} \phi^{mnp}  H_{mnp}  \qquad  H_{{\bf 7} \, m}  = \frac{1}{6} H^{npq} \psi_{npqm}  \qquad 
H_{{\bf 27} \, mn} =  \frac{1}{2} H_{ pq (m} \phi_{n)}^{\,\,\, pq} -  \frac{ 3}{7} H_{\bf 1} \delta_{mn} \, ,  \nonumber 
\eeq
and  comparing with the torsion of the $G_2$ structure 
\bseq
\begin{align}
\label{g2diff}
\d\varphi&=3\,W_{\mathbf 7}\wedge\varphi+W_{\mathbf 1}\psi+*W_{\mathbf{27}}\\
\d\psi&=4\,W_{\mathbf 7}\wedge\psi+*W_{\mathbf{14}} \, , 
\end{align}
\eseq
it is easy to show  that supersymmetry is not enough to completely determine the solution.  This can already be seen by simple counting 
of the full set  degrees of freedom in the solution and  the number of constraints imposed by supersymmetry, as summarised in the table below 
\begin{center}
\begin{tabular}{c|c|c}
$G_2$ repr.  &  parameters &  susy eq. \\
\hline 
$\mathbf{1}$ & $W_{\mathbf{1}}$ \,   $e^{-\Delta}\mu$ \,  $e^{-3\Delta}\tilde{H}$ \, $e^{-3\Delta}C\Lambda$  \, $H_{\mathbf{1}}$  \,  $\Lambda\Sigma_{\mathbf{1}}$  & 3 \\
$\mathbf{7}$ &$ W_{\mathbf{7}}$ \, $\d\phi $\, $\d\Delta$ \, $ H_{\mathbf{7}}$ \, $\Lambda\Sigma_{\mathbf{7}}$ & 3\\
$\mathbf{14}$ & $W_{\mathbf{14}}$ & 1 \\
$\mathbf{27}$ & $W_{\mathbf{27}}$ \, $ H_{\mathbf{27}}$  \, $\Lambda\Sigma_{\mathbf{27}}$  & 1
 \end{tabular} 
 \end{center}

More explicitly,  combining \eqref{eq:SUSY1} --  \eqref{eq:SUSY5} and \eqref{g2diff} we can fix  three of the singlets
\bseq
\begin{align}
& W_{\bf 1}  =  \frac{8}{7}  \mu e^{- \Delta} + \frac{1}{7}  e^{- 3 \Delta} ( 2 \tilde H  +  \a \Lambda C) \, ,  \\
& H_{\bf 1} = e^{-3\Delta} (2 \mu \, e^{2 \Delta} - \frac{\a}{2} \Lambda C )  \, , \\
&  \Lambda  \Sigma_{\bf 1} = -  \frac{4 i}{ \a}   e^{-3 \Delta} \, ( \tilde{H} + \frac{\a}{4} C \Lambda - 2 \mu e^{2 \Delta})  \, .
\end{align}
\eseq
Similarly, \eqref{eq:SUSY5},  \eqref{eq:SUSY3} with  \eqref{g2diff}   determine some of the forms in the ${\bf 7}$
\begin{align}
 W_{\bf 7}  =  \frac{1}{4} H_{\bf 7}   \qquad   \Lambda \Sigma_{\bf 7} =-  \frac{8 i }{  \a} \d \Delta
\end{align}
and  in the {\bf 27} 
\beq
W_{\bf 27} \sim   H_{\bf 27}  -  i\frac{\a}{2}\Lambda  \Sigma_{\bf 27} \, . 
\eeq
Finally notice that,   even with a generic condensate, it is not possible to generate the representation $\mathbf{14}$ of the torsion classes. $G_2$-structures of this type are often referred to as {\it integrable}. They have strong similarities with  even-dimensional complex manifolds \cite{fernandez1998dolbeault} and might 
be useful for 
a better  understanding of heterotic $G_2$-compactifications and their moduli (see e.g. \cite{delaOssa:2014lma, delaOssa:2016ivz} for work in this direction).

As in the previous section,  in order  to solve the equations of motion a solution of the supersymmetry variation has to satisfy the integrability constraints  \eqref{eq:Int2A} -- \eqref{eq:B}, imposing further constraints on $\Sigma$.

\subsection{Four-dimensional domain walls} 
\label{sec:dw4}

The  ansatz  \eqref{eq:compSeven} and  \eqref{eq:epsilon7split} also describes  $(3+1)$-dimensional domain walls. 
For these solutions the seven-dimensional metric decomposes as 
\begin{equation}
\d s_{7}^2=e^{2 A(x)} \d y^2+\d s_6^2\:,
\end{equation}
where $y$ is the normal coordinate to the domain wall and  $x^m$  are coordinates on the six-dimensional  internal manifold $M_6$.
In this paper we assume it to have   $SU(3)$ structure.
With this ansatz, the $G_2$-structure forms  can be written as
\bseq
\begin{align}
\label{eq:Firstdecomp1}
\varphi&=e^A\d y \wedge J+\Omega_-\\
\label{eq:Firstdecomp2}
\psi&=e^A\d y \wedge\Omega_+-\frac{1}{2}J\wedge J\:,
\end{align}
\eseq
where $\Omega=\Omega_++i\Omega_-$  and  $J$  are the  holomorphic three-form and the almost hermitian two-form defining  the  $SU(3)$ 
structure on $M_6$. 
The three-form flux $H$ and condensate three-form $\Sigma$  decompose as 
\bseq
\begin{align}
\label{dwH}
H&=\d y\wedge H_2+H_3\\
\Sigma&=\d y\wedge\Sigma_2+\Sigma_3\:,
\end{align}
\eseq
where $H_2$, $\Sigma_2$  and $H_3$  $\Sigma_3$ are two and three-forms on $M_6$. 

Plugging the ansatz above in  the system \eqref{eq:SUSY1}-\eqref{eq:SUSY3}, we find
\bseq
\begin{align}
& -\frac{1}{2}e^{-A }H_2\wedge J\wedge J+\Omega_+\wedge H_3 =* (2\mu e^{-\Delta}-\frac{\a}{2}C\Lambda e^{-3\Delta})\\
 & -i\frac{\a}{4} ( \frac{1}{2} e^{-A }\Lambda \Sigma_2 \wedge J\wedge J-\Omega_+\wedge \Lambda \Sigma_3 ) =* (e^{-3\Delta}\tilde H+\frac{\a}{4}e^{-3\Delta}C\Lambda-2\mu\, e^{-\Delta}) \, , 
\end{align}
\eseq
where $*$ denotes the six-dimensional Hodge star. The differential conditions become
\bseq
\begin{eqnarray}
 \d J &=&  -\d A\wedge J+e^{-A}\Omega_-'-(2\phi'-3\Delta')e^{-A}\Omega_-+(2\d\phi-3\d\Delta)\wedge J\notag\\
&  &  +* (H_3-i\frac{\a}{2}\Lambda\Sigma_3)-2\mu\, e^{-\Delta}\Omega_+\\
\d\Omega_- &= & (2\d\phi-3\d\Delta)\wedge\Omega_--* (H_2-i\frac{\a}{2}\Lambda\Sigma_2 )e^{-A}-\mu\,e^{-\Delta}J\wedge J\\
 \d \Omega_+  &=&  -e^{-A}  J\wedge J'   +  e^{-A}   (\phi'-\Delta')J\wedge J + (2 \d\phi- 2 \d\Delta - \d A)\wedge \Omega_+  \\ 
J\wedge\d J &= &   (\d\phi-\d\Delta)\wedge J\wedge J\\
 *\Delta' &=&  i\frac{\a}{8}e^{A}\Lambda \Sigma_3\wedge\Omega_-\\
 *\d\Delta &= & i\frac{\a}{8} (e^{-A}\Lambda \Sigma_2\wedge\Omega_--\Lambda \Sigma_3\wedge J ) \:,
\end{eqnarray}
\eseq
where the prime denotes derivatives with respect to the  $y$ direction. This system of equations should also be supplemented with the extra integrability constraints needed to solve the equations of motion, properly reduced to the domain wall situation. 
With the ansatz  \eqref{dwH} for $H$, the Bianchi identity splits into the two equations
\bseq
\begin{align}
\label{eq:DWfakeBI1}
\d \bar H_3&=\frac{\a}{4}\left(\tr\:F_2\wedge F_2-\tr\:R_2^-\wedge R_2^-\right)\\
\label{eq:DWfakeBI2}
\bar H'_{3}&=\d \bar H_2+\frac{\a}{2}\left(\tr\:F_1\wedge F_2-\tr\:R^-_{1}\wedge R^-_{2}\right)\:,
\end{align}
\eseq
where $\bar H = H -i  \frac{\a}{2} \Lambda \Sigma$ and we have defined
\bseq
\begin{align}
F&=F_2+\d y\wedge F_1\\
R^-&=R^-_{2}+\d y\wedge R^-_{1}\:.
\end{align}
\eseq

\section{Symplectic half-flat domain walls}
\label{sec:Ex}

Explicit solutions are not the focus of this paper, but we would like to reexamine 
an already known solution, which provides an illustration of the fake supersymmetry  approach of section \ref{sec:SpecialSol1}.   The idea is to look for geometries such that are solutions of the equations \eqref{fakesusy1}  and \eqref{fakesusy2}. 

We look for a domain wall solution as in section \ref{sec:dw4}.  The   ten-dimensional metric  is 
\begin{equation}
\label{eq:exmet}
\d s^2= \d s_3^2+   e^{2 A} \d y^2+\d s_6^2  \, , 
\end{equation}
where we set to zero the warp factor $\Delta$.\footnote{This follows from assuming that the
equations are formally the same as with no condensate.}  
The fake supersymmetry equations  are 
\begin{align}
\label{eq:DWfake1}
-\frac{1}{2}e^{-A}\bar H_2\wedge J\wedge J+\Omega_+\wedge\bar H_3=  2 \ast \mu  \:,\;\;\;\tilde{\bar H} =  2 \mu \:,
\end{align}
and the differential conditions 
\bseq
\begin{align}
\label{eq:DWfake2}
& J\wedge\d J=\:\d\phi\wedge J\wedge J\\
\label{eq:DWfake3}
&\d(e^A\Omega_+)= \:- J\wedge J'+\phi'J\wedge J+2\d\phi\wedge e^A\Omega_+\\
\label{eq:DWfake4}
& \d J=\:-\d A \wedge J+e^{-A}\Omega_-'-2\phi'e^{-A}\Omega_-+2\d\phi\wedge J+*\bar H_3 - 2 \mu \Omega_+ \\
\label{eq:DWfake5}
& \d\Omega_-= \:2\d\phi\wedge\Omega_--*\bar H_2e^{-A} - \mu J \wedge J \: , 
\end{align}
\eseq
where $\d$ now denotes the exterior derivative in six dimensions and  the  prime denotes the derivative along $y$.
The Bianchi identity imposes the  further  conditions \eqref{eq:DWfakeBI1} and \eqref{eq:DWfakeBI2}
\bseq
\begin{align}
\label{eq:DWfakeBI1b}
\d \bar H_3&=\frac{\a}{4}\left(\tr\:F_2\wedge F_2-\tr\:R_2^-\wedge R_2^-\right)\\
\label{eq:DWfakeBI2b}
\bar H'_{3}&=\d \bar H_2+\frac{\a}{2}\left(\tr\:F_1\wedge F_2-\tr\:R^-_{1}\wedge R^-_{2}\right) \, . 
\end{align}
\eseq
For zero condensate, this system has been analyzed to a great degree in the literature before, see e.g. \cite{Held:2010az, Lukas:2010mf, Klaput:2011mz, Gray:2012md, Klaput:2012vv, Klaput:2013nla, Maxfield:2014wea, Haupt:2014ufa}. Here we will see how it can also admit solutions with non trivial condensate.

Consider first  the system \eqref{eq:DWfake1}-\eqref{eq:DWfake5} together with the Bianchi identities \eqref{eq:DWfakeBI1b} and \eqref{eq:DWfakeBI2b}.
A relatively simple solution is obtained assuming that the internal six-dimensional geometry is an 
half-flat  manifold 
\bseq
\begin{align}
\label{halfflat}
& \d J =0\\
& \d\Omega_+ = 0  \\
& \d\Omega_-=W_2\wedge J\: , 
\end{align}
\eseq
where  $W_2$ is  purely imaginary and taking  the flux to be closed, i.e. we solve the Bianchi identity exactly.\footnote{
This can be achieved by embedding the gauge-connection in the $\nabla^-$ connection.} The Bianchi identity in this case reduces to
\begin{equation}
\label{eq:redBI}
\d \bar H_3=0\:,\;\;\;H'_{(3)}=\d \bar H_2\:,\;\;\; \d\mu=\mu'=0\:.
\end{equation}
In our example  the internal six-dimensional manifold is a particular solvmanifold  defined in \cite{tomassini2008symplectic}.  Let $G$ be the Lie group of matrices of the form
\begin{equation}
\left( \begin{array}{cccc}
e^{\lambda z} & 0 & 0 & \;\; x \\
0 & e^{-\lambda z} & 0 & \;\; y \\
0 & 0 & 1 & \;\;z \\
0 & 0 & 0 & \;\;1\end{array} \right)\:,
\end{equation}
where $\{x,y,z\}$ are real numbers and $\lambda=\log \frac{3+\sqrt{5}}{2}$. 
Then $G$ is a connected solvable Lie group admitting a cocompact lattice $\Gamma$,  so that the quotient  $N=G/\Gamma$ is a three-dimensional compact parallelizable solvmanifold \cite{fernandez1990compact}.   The six-dimensional solvmanifold is the product
 $M_0=N\times N$.  $M_0$ admits a coframe that  satisfies
\beq
\begin{array}{lcl}
\d\alpha_1=-\lambda\,\alpha_1\wedge\alpha_3    & \qquad   \qquad  &   \d\alpha_4 =-\lambda\,\alpha_4 \wedge\alpha_6  \\
\d\alpha_2=\lambda\,\alpha_2\wedge\alpha_3 & \qquad  \qquad   &    \d\alpha_5  = \lambda\,\alpha_5 \wedge\alpha_6 \\
\d\alpha_3=0 & \qquad  \qquad  &   \d\alpha_6   =0    \, \, 
\end{array}
\eeq
where  $\{\alpha_1,\alpha_2,\alpha_3\}$ is a coframe on $N$. 
It is easy to check that 
\bseq
\begin{align}
& J_0=\alpha_1\wedge\alpha_2+\alpha_4\wedge\alpha_5+\alpha_3\wedge\alpha_6 \, \\ 
& \Omega_0=\frac{(1-i)}{\sqrt{2}}\,(\alpha_1+i\,\alpha_2)\wedge(\alpha_4+i\,\alpha_5)\wedge(\alpha_3+i\,\alpha_6) \, 
\end{align} 
\eseq
define an half-flat structure. Moreover
\begin{align}
\Omega_{0+}&=\frac{1}{\sqrt{2}\lambda}\d P_0 
\end{align}
with $P_0=\alpha_{14}+\alpha_{15}-\alpha_{24}+\alpha_{25}$\footnote{We use the notation  $\alpha_{mn}=\alpha_m\wedge\alpha_n$.}
 is a primitive two-form
\begin{equation}
P_0\wedge J_0\wedge J_0=0\:.
\end{equation}

In our solution  we take the six-dimensional metric to be 
\beq
\d s_6^2 = e^{\phi(y)}  \d s_0^2
\eeq
where  $ \d s_0^2$ is  the metric on  the solvmanifold $M_0$ and, correspondingly, 
\beq
J = e^{\phi}  J_0 \qquad \Omega = e^{\frac{3 \phi}{2}}  \Omega_0 \, . 
\eeq 
Then  it is easy to see that \eqref{eq:DWfake2}  is solved if the 
dilaton only depends on the domain wall direction $\phi= \phi(y)$ so that
\beq
\d  \phi  =0 \, . 
\eeq
while  \eqref{eq:DWfake3} implies that the warp factor  must be constant  (for simplicity we take $A = 0$). 
\eqref{eq:DWfake4}  and  \eqref{eq:DWfake5} become 
\bseq
\begin{align}
\bar H_3  & =  -  \frac{1}{2}   e^{\frac{3 \phi}{2}}  ( \phi'   \Omega_{0+} -  4 \mu  \Omega_{0-} )  \\
*\bar H_2  &=  -   e^{\frac{3 \phi}{2}} \d  \Omega_{0-}  \, . 
\end{align}
\eseq
Then the Bianchi identity  $\d \bar{H}_3 = 0 $ in \eqref{eq:redBI}  together with  \eqref{halfflat}  imply that the three-dimensional space is Minkowksi,
$\mu =0$. Using the explicit form of $\Omega_0$ we find that  the fluxes  are
\bseq
\begin{align}
& \bar H_2=e^{\frac{\phi}{2}}\sqrt{2}\lambda\,P_0\\
& \bar H_3=-\frac{1}{2}\phi' e^{\frac{3 \phi}{2}}  \Omega_{0+}=-\frac{1}{3}\left(e^{\frac{3}{2}\phi}\right)'\Omega_{0+} \\ 
&\tilde{H}  = - \frac{\alpha^\prime}{2} C \Lambda
\end{align}
\eseq

We still have to impose the Bianchi identity $\bar{H}_3^\prime  = \d \bar{H}_2$, which reduces to a differential equation for the dilaton 
\begin{equation}
\label{eq:DilDiffRe}
\left(e^\phi\right)''+\frac{1}{2}\left(\phi'\right)^2\,e^\phi+4\lambda^2=0\:.
\end{equation}
From \eqref{eq:DilDiffRe}  we see that $e^{\phi}$ always has a negative second order derivative.
 There are two possibilities, either $e^\phi$ is bounded on an interval, or it tends to some linear function from below. The latter option is not possible however, as from \eqref{eq:DilDiffRe} this would imply that 
\begin{equation*}
\frac{1}{2}\left(\phi'\right)^2\,e^\phi+4\lambda^2\rightarrow0\: . 
\end{equation*}
By rescaling the $y$-coordinate, $\d y=e^{\frac{1}{2}\phi}\d t$, 
we find that the equation reduces to
\begin{equation}
\p_t^2 e^\phi +4\lambda^2\,e^\phi=0\:,
\end{equation}
with  solution
\begin{equation}
\label{eq:HFSol}
e^\phi=\,\big\vert\cos\big(2\lambda(t-t_0)\big)\big\vert\:,
\end{equation}
where we have set an overall constant to one, so that $\phi_0=\phi(t=t_0)=0$. 
Similar equations were discussed in \cite{delaOssa:2014lma}.
The solution \eqref{eq:HFSol} is periodic, and discontinuous at points where the cosine vanishes. The discontinuity leads to the following profile for $\d \bar H$, where now $\d=\d t\,\p_t+\d_6$
\begin{equation}
\label{eq:branesBI}
\d \bar H=-4\lambda^2\,\sum_n\delta\left[2\lambda(t-t_0)-\frac{\pi}{2}(1+2n)\right]\d t\wedge\Omega_{+0}\:,
\end{equation}
which suggest that there are sources localised  along the $t$-direction at intervals of $\Delta t=\pi/2\lambda$, while they wrap a trivial internal three-cycle, Poincar\'e dual to the exact form $\Omega_{+0}$. 

Let us also comment on the scalar curvature close to the source.  Write the seven-dimensional metric as
\begin{equation}
\d s_7^2=\d y^2+e^{\phi(y)}g_{0\,mn}\d x^m\d x^n=e^{\phi(t)}\left(\d t^2+g_{0\,mn}\d x^m\d x^n\right)\:.
\end{equation}
Consider the metric in the near-source region, i.e.
\begin{equation}
t-t_0=\frac{\pi}{4\lambda}+\Delta t\:,\;\;\;\Delta t>0\:,
\end{equation}
for small enough $\Delta t$. In this region the warp factor behaves like
\begin{equation}
e^\phi=\sin\left(2\lambda\Delta t\right)=2\lambda\Delta t+\OO(\Delta t^2)\:.
\end{equation}
With this, we find  the metric close to the source
\begin{equation}
\d s_7^2=\d y^2+\left(3\lambda y\right)^{2/3}\d s_0^2\:.
\end{equation}

Computing the Ricci scalar, we find
\begin{equation}
{\cal R}=\frac{1}{\sin\left(2\lambda\Delta t\right)}\left({\cal R}_0+24\lambda^2-6\cot(2\lambda\Delta t)\right)\:,
\end{equation}
where ${\cal R}_0$ is the Ricci scalar of $g_0$. This has the following pole expansion
\begin{equation}
{\cal R}=-\frac{3}{4 \lambda(\Delta t)^3}+\frac{{\cal R}_0+27\lambda^2}{2\lambda\Delta t}+{\cal O}(\Delta t)\:.
\end{equation}
Any half-flat symplectic solution of this form will have this behaviour. In particular, note that at some point close enough to the source the vacuum energy becomes negative. 

We conclude with two comments:
The solution of the system above is generically not supersymmetric. The additional constraint \eqref{eq:specialsol}, which leads to a supersymmetric solution,  in this case reduces to
\bseq
\begin{align}
\label{eq:DWfakeBPS1}
-\frac{1}{2}e^{-A}\Sigma_2\wedge J\wedge J+\Omega_+\wedge \Sigma_3&=-i*\,C\Lambda\\
\label{eq:DWfakeBPS2}
\Sigma_3\wedge\Omega_-&=0\\
\label{eq:DWfakeBPS3}
e^{-A}\Sigma_2\wedge\Omega_-&=\Sigma_3\wedge J\:.
\end{align}
\eseq
One may see that the ``canonical" form  of a seven-dimensional  condensate  $\Sigma\propto\varphi$ 
will not work here. For instance,  if we  consider condensates  of the form 
\begin{equation}
\Sigma=a \,\Omega_{0+}+b\,\Omega_{0-}+c\,\d y\wedge P_0+d\,\d y\wedge J_0\:,
\end{equation}
where  $a$, $b$, $c$ and $d$ are  $y$-dependent functions. From \eqref{eq:DWfakeBPS2} we see that we need $a=0$, while the  closure condition on $\Sigma$ gives
\begin{equation}
b\,\d\Omega_{0-}+b'\,\d y\wedge\Omega_{0-}-c\,\d y\wedge\d P_0=0\:,
\end{equation}
from which we get $b=c=0$. Hence, the only type of supersymmetry preserving solutions of this type are given by
\begin{equation}
\Sigma_3=0\:,\;\;\;\Sigma_2=d\,J_0\:.
\end{equation}
The value of $C\Lambda$ is the determined through \eqref{eq:DWfakeBPS1}.

Finally,  let us return to the sources in  \eqref{eq:branesBI}. They clearly are of codimension four, and one may be tempted to interpret  them as NS5-branes localised along $\d t$. It would be interesting to be able to confirm this and to study the  stability properties of the soluton.

\section*{Acknowledgments}

We would like to thank N. Halmagyi, C. Strickland-Constable, S. Theisen and  A. Tomasiello for helpful discussions.  RM  thanks the Korea Institute for Advanced Study and the Simons Center for Geometry and Physics, MP thanks Galileo Galilei Institute, and EES thanks the Theoretical Physics Department at CERN for hospitality during the course of the work. This work of RM was supported in part by the Agence Nationale de la Recherche under the grant 12-BS05-003-01. The work of EES, made within the Labex Ilp (reference Anr-10- Labx-63), was supported by French state funds managed by the Agence nationale de la recherche, as part of the programme Investissements dÕavenir under the reference Anr-11-Idex-0004-02.
\appendix

\section{Heterotic Supergravity}
\label{app:hetsugra}

The bosonic sector of heterotic supergravity consists of the metric $g$, the NS two-form $B$, the dilaton $\phi$ and the $E_8\times E_8$ gauge-field $A$. The corresponding fermionic fields are the graviton $\psi_M$, the dilatino $\lambda$ and the gaugino $\chi$. 
The action for the bosonic sector is
\begin{align}
\label{eq:Lagr0}
S_0=\int_{M_{10}}e^{-2\phi}\bigg[&\mathcal{R}+4\nabla_M\phi\nabla^M\phi-\frac{1}{12}H_{MNP}H^{MNP}\notag\\
&-\frac{\a}{8}\left(\tr\,F_{MN}F^{MN}-\tr\,R^-_{MN}{R^-}^{MN}\right)\bigg]\, , 
\end{align}
where the field strength $H$ is defined as
\begin{equation}
H=\d B+\frac{\a}{4}\Big(\omega_{CS}(A)-\omega_{CS}(\nabla^-)\Big) \, ,
\end{equation}
and satisfies  the Bianchi identity
\begin{equation}
\label{eq:BI}
\d H=\frac{\a}{4}\left(\tr\,F\wedge F-\tr\,R^-\wedge R^-\right)\:.
\end{equation}
The derivatives  $\nabla^\pm$ have connection symbols given by
\begin{equation}
\label{eq:connsymb}
{\Gamma^{\pm}_{KL}}^M={\Gamma^{\hbox{\tiny{LC}}}_{KL}}^M\pm\frac{1}{2}{\hat{H}_{KL}}^{\;\;\;\;M} \, . 
\end{equation}
The ten-dimensional bosonic equations of motion are 
\bseq
\begin{align}
\label{eq:eom1}
& \R_{MN}+2\nabla_M\nabla_N\phi-\frac{1}{2}H_{M}{H_N}  
-\frac{\a}{4}\Big(\tr\: F_{M}  \lrcorner {F_N}-\tr\: {R^-}_{M}  \lrcorner {R^-_N}\Big)=0+\OO(\a^2)\\
\label{eq:eom2}
& \frac{1}{4}\R+\nabla^2\phi-(\nabla\phi)^2-\frac{1}{8}H^2 
-\frac{\a}{16}\Big(\tr\,F^2 -\tr\, R^{- \, 2} \Big)=0+\OO(\a^2)\\
\label{eq:eom3}
& e^{2\phi}\nabla^M\Big(e^{-2\phi}H_{MNP}\Big)=0+\OO(\a^2) \\
\label{eq:eom4}
& e^{2\phi}\d_A(e^{-2\phi}*F)-F\wedge*H =0\:+\OO(\a) \, , 
\end{align}
\eseq
where the symbol $ \lrcorner$ denotes the contractions. In particular,  given a $p$-form $A_p$ and a $q$-form B, we defined
\bseq
\begin{eqnarray}
&& A_M \lrcorner  A_N = \frac{1}{(p-1)!} A_{M  M_1..M_{p-1}} A^{M M_1..M_{p-1}} \\
&& (A \lrcorner B)_{M_{p+1} \ldots M_q} = \frac{1}{p!} A^{N_1 \ldots N_p} B_{N_1 \ldots N_p M_{p+1} \ldots M_q}
\end{eqnarray}
\eseq

Since we are interested in solution respecting Poincar\'e invariance,  we will always  take the vacuum expectation value (vev) of individual fermions to vanish
\begin{equation}
\langle\psi_M\rangle=\langle\chi\rangle=\langle\lambda\rangle=0 \, , 
\end{equation}
so that the only relevant equations of motion are the bosonic ones.

The heterotic supersymmetry transformations with no fermionic bilinears are
\bseq 
\begin{align}
\label{eq:varpsiApp}
\delta\psi_M=\:&\nabla_M\epsilon+\frac{1}{8} H_{MNP}\Gamma^{NP}\epsilon  + \OO(\a^2)  \\
\label{eq:vardilApp}
\delta\lambda=\:&-\frac{\sqrt{2}}{4}\left[\slashed\p\phi+\frac{1}{2}\slashed  H\right]+\OO(\a^2)\\
\label{eq:varchiApp}
\delta\chi=\:&-\frac{1}{2\sqrt{2}}\slashed F \epsilon +\OO(\a^2)\:,
\end{align}
\eseq
where $\slashed A$ denote the Clifford product   
\begin{equation}
\slashed A= \frac{1}{p!} A_{M_1..M_p} \Gamma^{M_1..M_p} \:,
\end{equation}

The  connection $\Gamma^-$ in \eqref{eq:connsymb} is the bosonic part of an $SO(9,1)$ Yang-Mills gauge supermultiplet $({\Gamma^{-}_{KL}}^M,\psi_{PQ})$, 
whose fermionic part is the supercovariant curvature 
\begin{align}
\psi_{MN}=\:&\nabla^+_M\psi_N-\nabla^+_N\psi_M \, .
\end{align}
The fields $({\Gamma^{-}_{KL}}^M,\psi_{PQ})$ are constructed so as to transform as a $SO(9,1)$ Yang-Mills supermultiplet, modulo terms of $\OO(\a)$. Since these fields only appear in the theory at $\OO(\a)$, the theory is supersymmetric modulo terms  $\OO(\a^2)$. Note that this holds even when including the higher order fermionic terms in the action \cite{Bergshoeff1989439}.

\section{Conventions and G-structures in various dimensions}
\label{app:conv}
We Under the compactification
\begin{equation}
M_{10}=M_{10-d}\times X_d\:,
\end{equation}
we use the following index conventions for our coordinates
\begin{align}
\textrm{10-dimensional}&=I,J,K,..\notag\\
\textrm{$d$-dimensional}&=m,n,p,..\notag\\
\textrm{(10-$d$)-dimensional}&=\mu,\nu,\rho,..\notag
\end{align}
For frame indices we will use $A,B, ...$  in ten dimensions,   $\alpha, \beta, ...$  and $a,b,...$ for $10-d$ and  $d$-dimensional quantities, respectively. We also take a moment to collect some other conventions used. a $p$-form $\alpha$ is defined as
\begin{equation}
\alpha=\frac{1}{p!}\alpha_{m_1..m_p}\d x^{m_1..m_p}\:,
\end{equation}
with its Hodge dual given by
\begin{equation}
*\alpha=\frac{1}{(d-p)!\,p!}\epsilon_{n_1..n_{d-p}m_1..m_p}\alpha^{m_1..m_p}\d x^{n_1..n_{d-p}}\:.
\end{equation}
We denote the contraction of a $p$-form $\alpha$ by a $q$-form $\beta$ where $(p>q)$ as
\begin{equation}
\beta\lrcorner\alpha=\frac{1}{(p-q)!\,q!}\beta^{m_1..m_q}\alpha_{m_1..m_p}\d x^{m_{q+1}..m_p}\:.
\end{equation}
Note that
\begin{equation}
*(\alpha\wedge\beta)=\alpha\lrcorner*\beta\:.
\end{equation}
With this, in the case of compact geometries the adjoint of the exterior derivative and Laplacian read
\begin{equation}
\d^\dagger=*\d*\:,\;\;\;\Delta=\d*\d*+*\d*\d\:.
\end{equation}

\subsection{G-structures in different dimensions}
\label{app:Gstr}
In this appendix we summarise the basic features of the G-structures we need in this paper. These are  
$G_2$ structures in seven dimensions and Spin(7) structures in eight dimensions.

\subsection{$SU(3)$ structure in six dimensions}
\label{app:6d}
In compactification to four dimensions we consider the ten-dimensional metric takes the form
\begin{equation}
\d s_{10}^2=e^{2\Delta (y)}\d s_4^2+\d s_6^2 \, 
\end{equation}
where $\d s_6^2$ is the metric on a compact six-dimensional manifold $X_6$ and  $y$ denotes its  coordinates.
For the ten-dimensional gamma-matrices we take the following decomposition
\begin{align}
\label{six+fourgapp}
\Gamma^\mu&= e^{\Delta} \gamma^\mu\otimes\bf 1 \nonumber \\
\Gamma^m&=\gamma_{(4)} \otimes\gamma_m\:,
\end{align}
where $\mu = 0, \dots, 3$ and $m= 1, \ldots, 6$ and $\gamma_{(4)} =i\gamma^{0123}$ is the four-dimensional chiral gamma.
The six-dimensional gamma-matrices $\gamma_m$ are hermitian and purely imaginary and we define the six-dimensional
chiral gamma as 
\beq
\gamma_{(6)} = -i\gamma^{123456} \, .
\eeq
Then  ten-dimensional chiral gamma is 
\beq
\Gamma_{(10)} = \prod_{A = 0}^9 \Gamma_A  =   \gamma_{(4)} \otimes \gamma_{(6)} \, . 
\eeq

We decompose the ten-dimensional Majorana-Weil spinor $\epsilon$ as
\begin{equation}
\label{eq:spinordecomp4d}
\epsilon=e^A\zeta_+\otimes\eta_++e^A\zeta_-\otimes\eta_- \, ,
\end{equation}
where $\zeta_{\pm}$ are four-dimensional Weil spinors of positive and negative chirality, while $\eta_\pm$ are six-dimensional Weil spinors of opposite 
chirality\footnote{We choose the gamma matrices in such a way that   $(\eta_+)^\ast = \eta_-$.}
\begin{equation}
\gamma_4\zeta_{\pm}=\pm\zeta_\pm\:,\;\;\;\;\gamma\eta_\pm=\pm\eta_\pm \:.
\end{equation}

We assume that the spinor $\eta_+$ is globally defined on $X_6$ and hence defines an $SU(3)$-structure.
This is alternatively specified by the two globally-defined forms  
\bseq
\begin{align}
\label{Omegabil}
& \Omega_{mnp}=i\eta_-^\dagger\gamma_{mnp}\eta_+\\
\label{Jbil}
& J_{mn} =-i\eta_+^\dagger\gamma_{mn}\eta_+=i\eta_-^\dagger\gamma^{mn}\eta_- \, . 
\end{align}
\eseq
These satisfy the usual $SU(3)$ structure relations\footnote{Sometimes conventions are used where the volume form is $*1=\frac{1}{6}J^3$. This can be understood as a redefinition $J\rightarrow-J$.}
\begin{equation}
\label{eq:SU3rel}
*1=\frac{i}{8}\Omega\wedge\bar\Omega=-\frac{1}{6}J\wedge J\wedge J\:,\;\;\;\Omega\wedge J=0\:.
\end{equation}
We we also need the bilinears
\bseq
\begin{align}
& \eta_\pm^\dagger\gamma_{mnpq}\eta_\pm=-3\,J_{[mn}J_{pq]} \\
& \eta_\pm^\dagger \gamma_{mnprst}\eta_+=i \eta_\pm^\dagger \epsilon_{mnprst} \gamma_{(6)}\eta_+\:.
\end{align}
\eseq
We also have the following duality properties
\bseq
\begin{align}
*\Omega&=i\Omega\:,\;\;\;*\Omega_+=-\Omega_-\:,\;\;\;*\Omega_-=\Omega_+\\
*J&=-\frac{1}{2}J\wedge J \, . 
\end{align}
\eseq
The exterior derivatives of  the $J$ and $\Omega$ can be decomposed into irreducible representation of $SU(3)$
\bseq
\begin{align}
\d J&=\frac{3}{2}{\rm Im}(W_{1} \bar\Omega)+W_4\wedge J+W_3\\
\d\Omega &= W_1\,J\wedge J+W_2\wedge J+\bar W_5\wedge\Omega\:.
\end{align}
\eseq
where $W_1, W_{2}, W_{3}, W_{4}$ and $W_5$ are the $SU(3)$   torsion classes.
Recall that  $W_1$  is a complex singlet  zero-form, $W_2$  is a complex primitive two-form in the ${\bf 8} \oplus {\bf \bar{8}}$,  $W_3$  is a real primitive $(2,1)$ plus  $(1,2)$ form  in  $ {\bf 6}  \oplus  {\bf \bar{6}}$, $W_4$  is a real one-form in $ {\bf 3}  \oplus  {\bf \bar{3}}$, and $W_5$  is a complex one-form
in  $ {\bf 3}  \oplus  {\bf \bar{3}}$.

\subsection{$G_2$ structure in seven dimensions}
\label{app:G2str}

In this section we recall our conventions for compactifications to three dimensions
\begin{equation}
\label{eqapp:top7split}
M_{10}=M_3\times X_7\:,
\end{equation}
where $M_3$ is maximally symmetric three-dimensional spacetime, and $X$ is a seven-dimensional possibly non-compact internal space. The space \eqref{eqapp:top7split} is equipped with the metric
\begin{equation}
\d s^2=e^{2\Delta(y)}\d s_3^2+\d s_7^2\:.
\end{equation}
With this ansatz, the ten-dimensional gamma matrices  decompose as\footnote{Our conventions for the Pauli-matrices are
\begin{equation}
\sigma_1=\left( \begin{array}{cc}
0 & 1 \\
1 & 0\end{array} \right)\:,\;\;\;\;
\sigma_2=\left( \begin{array}{cc}
0 & -i \\
i & 0\end{array} \right)\:,\;\;\;\;
\sigma_3=\left( \begin{array}{cc}
1 & 0 \\
0 & -1\end{array} \right)\:.
\end{equation}} 
\begin{align}
\label{eqapp:gammasplit}
\Gamma_\mu& = e^{\Delta}  \gamma_\mu \otimes\sigma_3\otimes1\notag\\
\Gamma_m &= 1 \otimes \sigma_1 \otimes \gamma_m \:,
\end{align}
where $\mu = 0,1,2$ and $m= 1, \ldots, 7$.  The three-dimensional gamma-matrices $\gamma_\mu$ are real and the seven-dimensional matrices $\gamma_m$ are chosen purely imaginary. The matrices $\gamma_\mu$ and $\gamma_m$ satisfy 
\beq
\label{eqapp:gammaprod} 
\gamma_{\mu \nu \rho} =  \epsilon_{ \mu \nu \rho}  \,  \mathbb{I}_{(2)} \qquad  \gamma_{1 \dots 7} =   i  \mathbb{I}_{(8)} 
\eeq
with $ \epsilon_{0 1 2} =1$ and $\epsilon^{012}  = - 1$. 
Then ten-dimensional  chirality operator is $\Gamma_{(10)}  =  -  \mathbb{I} \otimes\sigma_2 \otimes \mathbb{I}_{(8)}$. 

The supersymmetry parameter  $\epsilon$ has positive chirality in 10 dimensions and  splits as 
\begin{equation}
\label{eqapp:epsilon7splitapp}
\epsilon= e^{A(y)}   \zeta \otimes\left( \begin{array}{c}1 \\ - i \end{array} \right)\otimes\eta \, , 
\end{equation}
where $\eta$ is a a globally defined spinor of unitary norm  defining a $G_2$-structure on $X_7$. 

The $G_2$-structure  can be also defined in terms of bilinears  of $\eta$, namely 
a  three-form $\varphi$ and dual four-form $\psi=*\varphi$ which are $G_2$ singlets. 
If we take the spinor $\eta$ to have norm one
$\varphi$ and $\psi$ are given by 
\begin{equation}
\label{eqapp:defphipsi}
\varphi_{mnp}=-i \eta^\dagger\gamma_{mnp} \eta\:,\;\;\;\psi_{mnpq} = \eta^\dagger\gamma_{mnpq} \eta\, . 
\end{equation}

The exterior derivatives of  the $G_2$ invariant forms can be decomposed into irreducible representation of $G_2$
\bseq
\begin{align}
\d\varphi&=3\,W_{\mathbf 7}\wedge\varphi+W_{\mathbf 1}\psi+*W_{\mathbf{27}}\\
\d\psi&=4\,W_{\mathbf 7}\wedge\psi+*W_{\mathbf{14}}\:.
\end{align}
\eseq
where $\{W_{\mathbf 1}, W_{\mathbf 7}, W_{\mathbf{14}}, W_{\mathbf{27}} \}$ 
are   the four $G_2$  torsion classes  and the boldface numbers denote the various $G_2$ representations.
In particular  $W_{\mathbf 1}$ is a real scalar,  $W_{\mathbf 7}$ a real vector,  $W_{\mathbf{14}}$ 
and $W_{\mathbf{27}} $ is a symmetric, traceless tensor.

\subsection{Spin(7)  structure in eight  dimensions}
In eight-dimension a globally defined spinor $\eta$ gives rise to a $Spin(7)$ structure, whose fundamental four-form is given by
\begin{equation}
\label{spin7ff}
\Phi_{\alpha\beta\gamma\rho}=\eta^\dagger\gamma_{\alpha\beta\gamma\rho}\eta\:.
\end{equation}
The 16-dimensional spinors split into positive and negative chirality ones. 
If we choose $\eta$ to be of negative chirality,  a basis for the space of positive  chirality spinors is given by $\gamma_\alpha \eta$, while for 
the negative chirality ones we have 
\begin{equation}
\label{8dspinorbasis}
\eta \, , \qquad  \Pi^{\alpha \beta}_{{\bf 7} \, \, \gamma \delta}  \hat \gamma^{\gamma \delta} \eta \, .
\end{equation}
$ \Pi^{\alpha \beta}_{{\bf 7}  \, \, \gamma \delta}$ is the projector onto the representation ${\bf 7}$ and is given by 
\begin{equation}
\label{eq:Pi7}
 \Pi_{{\bf 7} \alpha \beta  \gamma \delta}  =  \frac{1}{8} ( g_{\alpha \gamma}  g_{\beta \delta} -g_{\alpha \delta}  g_{\beta \gamma} - \Phi_{ \alpha \beta  \gamma \delta})  \, . 
\end{equation}
Using the fundamental form  one can decompose the action of all other gamma matrices in terms of the spinorial basis   \cite{Marino:1999af}
\bseq
\begin{align}
\hat \gamma_{\alpha\beta}\eta&=-\frac{1}{6}\Phi_{\alpha\beta\gamma\rho} \hat \gamma^{\gamma\rho}\eta \\ 
\hat \gamma_{\alpha\beta\gamma}\eta&=\Phi_{\rho\alpha\beta\gamma} \hat \gamma^\rho\eta\\
\hat \gamma_{\alpha\beta\gamma\rho}\eta&=\Phi_{\alpha\beta\gamma\rho}\eta+{\Phi_{[\alpha\beta\gamma}}^\kappa \hat \gamma_{\rho]\kappa}\eta\\
\hat \gamma_{\alpha\beta\gamma\rho\kappa}\eta&=5\,\Phi_{[\alpha\beta\gamma\rho} \hat \gamma_{\kappa]}\eta\:.
\end{align}
\eseq
We will need to decompose Spin(7) into $G_2$.  The eight-dimensional gamma matrices $\hat \gamma_\alpha$ are given in terms of the seven-dimensional ones
by
\begin{equation}
\hat \gamma^i = \sigma_2 \otimes \gamma^i  \qquad \qquad  \hat \gamma^8 = - i \sigma_1 \otimes \gamma^8 = \sigma_1 \otimes 1  \, ,
\end{equation}
with $i=1, \dots, 7$ and where we defined $\gamma^8 = - \gamma^1  \ldots \gamma^7$. Moreover, the fundamental four-form decomposes as
\begin{equation}
\Phi=\varphi\wedge\d x^{8}+\psi\:,
\end{equation}
where $\varphi$ is the fundamental three-form specifying the $G_2$-structure, and $x^{8}$ is the eight-dimensional special direction.

\section{Derivation of the supersymmetry conditions for $G_2$ structures in seven dimensions}
\label{app:deriv}
In this section we discuss how to derive the set of equations \eqref{eq:SUSY1}-\eqref{eq:SUSY3}. 
For simplicity of notation we define the combination
\beq
H^\pm = H \pm \frac{\a}{4} \<\Sigma \>
\eeq
whose non-trivial components are
\bseq
\begin{align}
\label{eqapp:Hsplit}
& H^\pm_{mnp} = H_{mnp} \mp  \frac{i}{4} \alpha^\prime  \Lambda \Sigma_{mnp} \\ 
& \tilde{H}^\pm_{\mu \nu \rho}  = \epsilon_{\mu \nu \rho}  \tilde{H}^\pm =   \epsilon_{\mu \nu \rho}  (\tilde H \pm \frac{\alpha^\prime}{4}  \Lambda  C) \:.
\end{align}
\eseq
Let us consider first the dilatino equation \eqref{eq:vardil}. 
Splitting into  three- and seven-dimensional indices
\begin{equation}
\label{eq:7Ddilationo1}
\Gamma^m \p_m \phi\epsilon+\frac{1}{2}H^-_{mnp}\Gamma^{mnp}\epsilon+\frac{1}{12}H^-_{\mu \nu \rho}\Gamma^{\mu \nu \rho }\epsilon=0 \, , 
\end{equation}
using the decompositions \eqref{eqapp:gammasplit},  \eqref{eqapp:epsilon7splitapp} and 
\eqref{eqapp:gammaprod}, we can reduce it to  an equation on the internal space only
\begin{equation}
\label{eq:7Ddilationo2}
\left(\slashed\p\phi+\frac{1}{2}\slashed H^--\frac{i}{2}  e^{- 3 \Delta} \tilde H^- \right)\eta=0 \, . 
\end{equation}

Consider now  the gravitino variation \eqref{eq:varpsi}. Choosing the frame
\begin{equation}
e_M^{\,\, A} =  \left\{ e^\Delta {e_\mu}^{\alpha}(x)\:,{e_m}^{a}(y)\right\} 
\end{equation}
for the metric \eqref{eq:compSeven},  it is straightforward to see that the spin-connection\footnote{We recall  that 
the ten-dimensional Levi-Civita equation for a generic spinor $\epsilon$ is given by
\begin{equation}
\nabla_M\epsilon=\p_M\epsilon+\frac{1}{4}{\omega_{M}}^{AB}\Gamma_{AB}\,\epsilon\:,
\end{equation}
where $\{A,B,..\}$ are  ten-dimensional frame-indices.} 
\begin{align}
{\omega_M}^{AB}&=\frac{1}{2}e^{N\,A}\left(\p_M e_N^B-\p_N e_M^B\right)-\frac{1}{2}e^{N\,B}\left(\p_M e_N^A-\p_N e_M^A\right)\notag\\
&-\frac{1}{2}e^{P\,A}e^{Q\,B}\left(\p_P e_{Q\,C}-\p_Q e_{P\,C}\right)e_M^C\:,
\end{align}
has non-zero components 
\begin{equation}
\label{spinconsplit}
{\omega_\mu}^{\alpha \beta}\:,\;\;\;{\omega_\mu}^{\alpha b}=e^\Delta{e_\mu}^{\alpha}e^{n b }\p_n\Delta\:,\;\;\;{\omega_m}^{a b}\:.
\end{equation}
Then, using \eqref{Sigmaans},   the $M=\mu$ component of the gravitino variation reduces to 
\beq
\label{eq:7Dgrav2}
\delta\psi_\mu =\nabla_\mu\epsilon +\frac{1}{2} \partial_n \Delta \Gamma_\mu \Gamma_n \epsilon 
+\frac{1}{8}H^+_{\mu \nu \rho}\Gamma^{\nu \rho}\epsilon - i  \frac{\a}{96} \Lambda \Sigma_{mnp} \Gamma^{mnp}\Gamma_\mu \epsilon=0 \, , 
\eeq
where we used the properties of gamma matrices to reconstruct the tensor  $H^+_{\mu \nu \rho}$ 
given in \eqref{eqapp:Hsplit}. The three-dimensional covariant derivative acts as 
\begin{equation}
\label{3dcov}
\nabla^{(3)}_\mu \zeta=\frac{\mu}{2}\gamma_\mu \zeta\:.
\end{equation}
where $- |\mu|^2$ the Anti-de-Sitter scalar curvature.   Plugging \eqref{3dcov} 
 \eqref{eqapp:gammasplit},  \eqref{eqapp:epsilon7splitapp} in \eqref{eq:7Dgrav2}, we find again a purely internal equation 
\begin{equation}
\label{eq:extgrav}
\left(\slashed\p\Delta + i\mu\,e^{-\Delta}-\frac{i}{2}e^{-3\Delta}\tilde H^+ +i \frac{\a}{8} \Lambda \slashed\Sigma\right)\eta=0\: .
\end{equation}
Similarly the $M=m$ component of the gravitino equation  can be written as 
\beq
\label{eq:7Dgrav1}
\delta\psi_m=\nabla_m\epsilon+\frac{1}{8}H^+_{mnp}\Gamma^{np}\epsilon - i  \frac{\a}{96}  \Lambda \Sigma_{npq} {\Gamma^{npq}}_m\epsilon + \frac{\a}{96}  
\<\Sigma_{\mu \nu \rho}\> \Gamma^{\mu \nu \rho}\Gamma_m\epsilon=0 \, ,
\eeq
and  using \eqref{eqapp:gammasplit} and  \eqref{eqapp:epsilon7splitapp} together with \eqref{Sigmaans} we find
 \begin{equation}
\label{eq:intGrav}
\nabla_m \eta+\p_m A+\frac{1}{8}H^+_{mnp}\gamma^{np}\eta -i \frac{\a}{96} \Lambda \Sigma_{npq}{\gamma^{npq}}_m\eta + i \frac{\a}{16}  C \Lambda \gamma_m\eta=0\:.
\end{equation}
Notice that combining \eqref{eq:intGrav} with the condition that  $\eta$ has unit norm we find
\beq
\label{eqapp:Acond}
0 = \nabla_m (\eta^\dagger \eta) = - 2 \partial_m A + i \frac{\a}{48} \Lambda \,  \Sigma_{npq} \psi^{npq}_{\quad m}  \, , 
\eeq
where $\psi$ is the $G_2$ structure four-form \eqref{eqapp:defphipsi}, and we used the fact that $\Sigma_{mnp}$ is purely imaginary while $\Lambda$ is
real.

It is convenient to rewrite the equations \eqref{eq:7Ddilationo2}, \eqref{eq:extgrav} and \eqref{eq:intGrav} as a set of conditions on the forms $\varphi$ and $\psi$ defining the $G_2$ structure.  The procedure is standard:  multiplying  each equation by $\eta^\dagger \gamma^m$, $\eta^\dagger \gamma^{mn}$ up to $\eta^\dagger \gamma^{mnpq}$ gives a set of equations for $\varphi$ and $\psi$. For the dilatino equation \eqref{eq:7Ddilationo2} we obtain the following set of equations 
\bseq
\begin{align}
\label{eq:dil1}
& e^{-3\Delta}\tilde H^-=\frac{1}{3!}H^-_{mnp}\varphi^{mnp}\\
\label{eq:dil2}
& \p_m\phi\,=\frac{1}{12}H^-_{npq}{\psi^{npq}}_m\\
\label{eq:dil3}
& \p_{[q}\phi\,\psi_{rstu]} =H^-_{m[qr}\,{\psi_{stu]}}^m\\
\label{eq:dil4}
& 4\p_{[m}\phi\,\varphi_{npr]} =\frac{1}{2}\,\left(*H^-\right)_{mnpr}-3\,H^-_{w[mn}\,{\varphi_{pr]}}^w-\frac{1}{2}e^{-3\Delta}\tilde H^-\psi_{mnpr}\:.
\end{align}
\eseq
The last two equations are not independent, they can be obtained from the first two using the properties of the $G_2$ structure forms. 
We give them since we will need them to simply expressions later on. More importantly, it should be stressed that that these equations contain only information about the $\mathbf{1}$ and $\mathbf{7}$ representations of the $G_2$-structure.

The  external gravitino \eqref{eq:extgrav} gives 
\bseq
\begin{eqnarray}
\label{eq:extgrav1}
&& \left(- 2\mu\,e^{-\Delta}+e^{-3\Delta}\tilde H^+\right) =  i \frac{\a}{24} \Lambda \Sigma_{mnp}\varphi^{mnp} \\
\label{eq:extgrav2} 
&&  \p_m\Delta =   i \frac{\a}{48} \Lambda \Sigma_{npq}{\psi^{npq}}_m \\
\label{eq:extgrav3}
&& \p_{[m}\Delta\,\psi_{npqr]}=\; i\frac{\a}{4} \Lambda  \Sigma_{s[mn}\,{\psi_{pqr]}}^s \\
\label{eq:extgrav4}
&&  4 \p_{[m} \Delta\,\varphi_{npq]}= \; - i\frac{3}{4}\a\Lambda \,\Sigma_{r[mn}\,{\varphi_{pq]}}^r + i \frac{\a}{8} \Lambda \,\left(*\Sigma\right)_{mnpq}  \notag \\
&& \qquad  \qquad \qquad  - \frac{1}{2} \left(e^{-3\Delta}\tilde H^+ - 2\mu\,e^{-\Delta}\right)\psi_{mnpq}\:.
\end{eqnarray}
\eseq
As before, the  last two equations are redundant, but we include them since we will use them to simplify later expressions. 

From \eqref{eq:extgrav2} it follows immediately that, in order to allow for a non-zero condensate, the warp factor need not be constant 
$\p_m\Delta\neq0$.  Moreover,  combining \eqref{eq:extgrav2}  and  \eqref{eqapp:Acond} we find 
\begin{equation} 
\d A= \frac{1}{2}  \d\Delta\:.
\end{equation}

Using equation \eqref{eq:intGrav}, together with the dilatino equations \eqref{eq:dil1}-\eqref{eq:dil4}, and the external gravitino \eqref{eq:extgrav1}-\eqref{eq:extgrav4} we find that 
\bseq
\begin{align}
\label{eq:dvarphiapp}
\d\varphi&=\left(2\d\phi-3\d\Delta\right)\wedge\varphi- ( \ast H-i\frac{\a}{2} \ast (\Lambda  \Sigma)) + (2\mu  \,e^{-\Delta} )  \,\psi\\
\label{eq:dpsiapp}
\d\psi&=\left(2\d\phi-2\d\Delta\right)\wedge\psi\:.
\end{align}
\eseq

\section{Derivation of the supersymmetry conditions for $SU(3)$-structures in six dimensions}
\label{app:deriv4d}
In this section we  derive the set of equations \eqref{eq:4dBPS1}-\eqref{eq:4dBPS5}. 
We use the  conventions appendix \ref{app:6d} for gamma matrices, spinors and SU(3) structure.  
Splitting into four and six-dimensional indices the supersymmetry variations \eqref{eq:varpsi}-\eqref{eq:varchi}  reduce to
\bseq
\begin{align}
\label{eq:4dvarPsiExt}
\nabla_\mu\epsilon+\frac{1}{2}\Gamma_\mu\partial_m\Delta\Gamma^m\epsilon+\frac{\a}{96}\langle\Sigma\rangle_{mnp}\Gamma^{mnp}\Gamma_\mu\epsilon&=0\\
\label{eq:4dvarPsiInt}
\nabla_m\epsilon+\frac{1}{8}H_{mnp}\Gamma^{np}\epsilon+\frac{\a}{96}\langle\Sigma\rangle_{rst}\Gamma^{rst}\Gamma_m\epsilon&=0\\
\label{eq:4dvarPhi}
\left(\Gamma^m\p_m\phi+\frac{1}{12}H_{mnp}\Gamma^{mnp}-\frac{\a}{48}\langle\Sigma\rangle_{mnp}\Gamma^{mnp}\right)\epsilon&=0\:.
\end{align}
\eseq

The only non-trivial components of the three-form flux $H$ and the gaugino three-form $\Sigma$ are purely internal and we define
\begin{equation}
H_{mnp}^\pm=H_{mnp}\pm\frac{\a}{4}\langle\Sigma_{mnp} \rangle\:.
\end{equation}

Consider first the external gravitino  \eqref{eq:4dvarPsiExt}. 
The ten-dimensional spin connection decomposes as in \eqref{spinconsplit} and the 
 four-dimensional  covariant derivative is
\begin{equation}
\label{nabla4}
\nabla^{(4)}_\mu\zeta_+=\frac{\mu}{2}\gamma_\mu\zeta_-\: , 
\end{equation}
where $\mu$ is a complex parameter related and the cosmological constant is equal to $- |\mu|^2$.  $\zeta_+$ is the four-dimensional susy parameter in
\eqref{eq:spinordecomp4d}. Using the splittings  \eqref{six+fourgapp} the equation reduces to 
\begin{equation}
\label{eq:4dExtGrav0}
\mu\,e^{-\Delta}\,\eta_+-\slashed\p\Delta\eta_-+\frac{\a}{8}\langle\slashed\Sigma\rangle\eta_-=0\:.
\end{equation}
Multiplying \eqref{eq:4dExtGrav0} by  $\eta_+^\dagger$ we find
\begin{equation}
\label{eq:4dExtGrav1}
\mu e^{-\Delta}=i\frac{\a}{8}\langle\Sigma\rangle\lrcorner\bar\Omega\: , 
\end{equation}
while multiplying by  $\eta_-^\dagger\gamma_r$  and taking the real part gives 
\begin{equation}
\label{eq:dDelta4d}
\p_m\Delta=\frac{\a}{16}\langle\Sigma\rangle^{rst}J_{[rs}J_{tm]}\: , 
\end{equation}
which can also be written as
\begin{equation}
\label{eq:4dBPS2A}
\frac{\a}{4}* \<\Sigma \> \wedge J=\d\Delta\wedge J\wedge J\:. 
\end{equation}
This exhausts the set of independent equations that can be derived from \eqref{eq:4dvarPsiExt}.
It is however useful to consider the equation obtained multiplying   \eqref{eq:4dExtGrav0}
by  $\eta_-^\dagger\gamma_{mnp}$ (and taking the real real part) and by  $\eta_+^\dagger\gamma_{mnpq}$  
\bseq
 \begin{align}
 \label{extgextra1}
\frac{3\a}{8}{\langle\Sigma\rangle^t}_{[mn}J_{p]t}&=   \frac{\a}{8}\left(*\langle\Sigma\rangle\right)_{mnp}+e^{-\Delta}\,{\rm Re}\left(\mu\,\Omega_{mnp}\right)-3\p_{[m}\Delta\,J_{np]} \, , \\
\label{extgextra2}
-\frac{3\a}{8}{\langle\Sigma\rangle^s}_{[mn}\Omega_{pq]s}&=-2\p_{[m}\Delta\,\Omega_{npq]}-i\frac{3}{2}e^{-\Delta}\bar\mu\,J_{[mn}J_{pq]}\ \, . 
\end{align}
\eseq
 With the splitting  \eqref{six+fourgapp} and  \eqref{eq:spinordecomp4d}   the dilatino equation  \eqref{eq:4dvarPhi}  becomes 
\begin{equation}
\label{eq:4dDil0}
\left(\slashed\p\phi+\frac{1}{2}\slashed H^-\right)\eta_+=0 \, . 
\end{equation}
As for the external gravitino, we can derive form equations by multiplying by the basis spinors 
$\eta_-^\dagger$, $\eta_+^\dagger \gamma_m$
\bseq
\begin{align}
\label{eq:4dDil1}
& H^-\lrcorner \Omega=  0 \, , \\ 
&\p_m\phi=-\frac{1}{4}{H^-}^{rst}J_{[rs}J_{tm]} \, .
\end{align}
\eseq
The second equation can also be written as
\begin{equation}
\label{eq:4dpreDil}
{*H^-}\wedge J=-J\wedge J\wedge\d\phi\:.
\end{equation}
Moreover multiplying \eqref{eq:4dDil0} by $\eta_+^\dagger \gamma_{mnp}$ and $\eta_-^\dagger \gamma_{mnpq}$ we find the two additional equations
\bseq
\begin{align} 
 \label{dilextra1}
 {{H^-}^t}_{[mn}J_{p]t} &=2\p_{[m}\phi\,J_{np]}+\frac{1}{3}\left(*H^-\right)_{mnp} \, , \\
\label{dilextra2}
\frac{3}{2}{{H^-}^s}_{[mn}\Omega_{pq]s}&=-2\p_{[m}\phi\,\Omega_{npq]}  \, . 
\end{align} 
\eseq
We now turn to the internal gravitino equation, which reduces to 
\begin{equation}
\label{extgspl}
\nabla_m\eta_++\p_mA\eta_++\frac{1}{8}H^+_{mnp}\gamma^{np}\eta_++\frac{\a}{96}\langle\Sigma\rangle_{rst}{\gamma^{rst}}_m\eta_+=0\:.
\end{equation}
As in the seven dimensional case, we insist that the internal spinors are normalised unit norm, from which it follows that
\begin{equation}
0=\nabla_m\left(\eta_+^\dagger\eta_+\right)=-2\p_mA + \frac{\a}{16}\langle\Sigma\rangle_{rst}J^{rs}{J^t}_m\:.
\end{equation}
Using this and \eqref{eq:dDelta4d}, we find
\begin{equation}
\d\Delta=2\d A\: .
\end{equation}

We can use \eqref{extgspl}  and the definitions  \eqref{Jbil} and  \eqref{Omegabil}
to compute the exterior derivatives of $J$ and $\Omega$.
For the $J$   we find 
\begin{align}
\nabla_{[p}J_{mn]}
\label{eq:4dpredJ}
&=-\p_{[p}\Delta\,J_{mn]}+{H^+}^t_{[mn}J_{p]t}-\frac{\a}{8}{\langle\Sigma\rangle}^t_{[mn}J_{p]t} + \frac{\a}{8}  ( \ast \langle\Sigma\rangle)_{mnp} \:.
\end{align}
Using  \eqref{extgextra1}  and  \eqref{dilextra1},  \eqref{eq:4dpredJ} reduces to 
\begin{equation}
\label{eq:4dpredJ1}
\d J=(2\d\phi-4\d\Delta)\wedge J+*\bar H+3\,e^{-\Delta}{\rm Re}\left(\mu\,\Omega\right)\:,
\end{equation}
where $\bar H=H+\frac{\a}{2}\langle\Sigma\rangle$

We next compute the exterior derivative of $\Omega$ to find
\begin{align}
\nabla_{[m}\Omega_{npq]}
\label{eq:4dpredOm}
&=-\p_{[m}\Delta\,\Omega_{npq]}-\frac{3}{2}{{H^-}^s}_{[mn}\Omega_{pq]s}-\frac{3\a}{8}{\langle\Sigma\rangle^s}_{[mn}\Omega_{pq]s} \, , 
\end{align}
which using  \eqref{extgextra2}  and  \eqref{dilextra2} becomes 
\begin{equation}
\label{eq:4dpredOm1}
\d\Omega=(2\d\phi-3\d\Delta)\wedge\Omega-ie^{-\Delta}\,\bar\mu\,J\wedge J\:.
\end{equation}

Note that equations \eqref{eq:4dDil1} can be derived by contracting \eqref{eq:4dpredJ1} with $\Omega$ and using \eqref{eq:4dExtGrav1} and \eqref{eq:4dpredOm1}. Furthermore, by wedging \eqref{eq:4dpredJ1} with $J$ and using \eqref{eq:4dBPS2A} and \eqref{eq:4dpreDil} we find
\begin{equation}
J\wedge\d J=(\d\phi-\d\Delta)\wedge J\wedge J\:.
\end{equation}
To summarise, the set of indepentent four dimensional BPS equations are
\begin{align}
\label{eq:4dBPS10}
\mu e^{-\Delta}=i\frac{\a}{8}\langle\Sigma\rangle\lrcorner\bar\Omega\:,
\end{align}
with the differential conditions
\bseq
\begin{align}
\label{eq:4dBPS20}
\frac{\a}{4}* \<\Sigma \>\wedge J&=\d\Delta\wedge J\wedge J\\
\label{eq:4dBPS30}
J\wedge\d J&=(\d\phi-\d\Delta)\wedge J\wedge J\\
\label{eq:4dBPS40}
\d J&=\left(2\d\phi-4\d\Delta\right)\wedge J+*\bar H+3\,e^{-\Delta}{\rm Re}\left(\mu\,\Omega\right)\\
\label{eq:4dBPS50}
\d\Omega&=\left(2\d\phi-3\d\Delta\right)\wedge\Omega-ie^{-\Delta}\,\bar\mu\,J\wedge J\:.
\end{align}
\eseq

\providecommand{\href}[2]{#2}\begingroup\raggedright\endgroup

\end{document}